\documentclass
[aps,floats,twoside,fancyhdr,superscriptaddress,nofootinbib,showpacs]{revtex4}%
\usepackage{latexsym}
\usepackage{amsmath}
\usepackage{amssymb}
\usepackage{amsfonts}
\usepackage{graphicx}
\usepackage{dcolumn}
\usepackage{fancyhdr}
\usepackage[colorlinks]{hyperref}%
\providecommand{\U}[1]{\protect\rule{.1in}{.1in}}
\newtheorem{theorem}{Theorem}

\newtheorem{remark}[theorem]{Remark}

\begin{document}
\title[Bianchi I]{About Bianchi I with VSL.}
\author{Jos\'{e} Antonio Belinch\'{o}n}
\affiliation{Dept. of Physics. ETS Architecture. UPM Av. Juan de Herrera 4. Madrid 28040. Espa\~{n}a.}
\email{abelcal@ciccp.es}
\date{\today }
\pacs{PACS number}

\begin{abstract}
In this paper we study how to attack, through different techniques, a perfect
fluid Bianchi I model with variable $G,c$ and $\Lambda,$ \textquotedblleft
but\textquotedblright\ taking into account the effects of a \textquotedblleft%
$c$-variable\textquotedblright\ into the curvature tensor. We study the model
under the assumption, $div(T)=0.$ These tactics are: Lie groups method (LM),
imposing a particular symmetry, self-similarity (SS), matter collineations
(MC) and kinematical self-similarity (KSS). We compare both tactics since they
are quite similar (symmetry principles). We arrive to the conclusion that the
LM is too restrictive and brings us to get only the flat FRW solution. The SS,
MC and KSS approaches bring us to obtain all the quantities depending on
$\left(  \int c(t)dt\right)  .$ Therefore, in order to study their behavior we
impose some physical restrictions like for example the condition $q<0$
(accelerating universe). In this way we find that $c$ is a growing time
function and $\Lambda$ is a decreasing time function whose sing depends on the
equation of state $\omega,$ while the exponents of the scale factor must
satisfy the conditions $\sum_{i=1}^{3}\alpha_{i}=1$ and $\sum_{i=1}^{3}%
\alpha_{i}^{2}<1,$ $\forall\omega$, i.e. for all equation of state$,$ relaxing
in this way the Kasner conditions. The behavior of $G$ depends on two
parameters, the equation of state $\omega$ and $\varepsilon,$ a parameter that
controls the behavior of $c(t),$ therefore $G$ may be growing or decreasing$.$
We also show that through the Lie method, there is no difference between to
study the field equations under the assumption of a $c-$var affecting to the
curvature tensor which the other one where it is not considered such effects.
Nevertheless, it is essential to consider such effects in the cases studied
under the SS, MC, and KSS hypotheses.

\end{abstract}
\maketitle


\section{Introduction.\label{intro}}

Since the pioneering work of Dirac (\cite{1}), who proposed, motivated by the
occurrence of large numbers in Universe, a theory with a time variable
gravitational coupling constant $G$, cosmological models with variable $G$ and
nonvanishing and variable cosmological term, $\Lambda,$ have been intensively
investigated in the physical literature (see for example \cite{2}-\cite{14})
and for alternative approaches in this framework see for example (\cite{15})
where the authors study a FRW model with variable equation of state and
(\cite{16}) with variable deceleration parameter, finding a positive
decreasing cosmological \textquotedblleft constant\textquotedblright.

In modern cosmological theories, the cosmological constant remains a focal
point of interest (see \cite{cc1}-\cite{cc4}\ \ for reviews of the problem). A
wide range of observations now compellingly suggest that the universe
possesses a non-zero cosmological constant. Some of the recent discussions on
the cosmological constant \textquotedblleft problem\textquotedblright\ and on
cosmology with a time-varying cosmological constant point out that in the
absence of any interaction with matter or radiation, the cosmological constant
remains a \textquotedblleft constant\textquotedblright. However, in the
presence of interactions with matter or radiation, a solution of Einstein
equations and the assumed equation of covariant conservation of stress-energy
with a time-varying $\Lambda$ can be found. This entails that energy has to be
conserved by a decrease in the energy density of the vacuum component followed
by a corresponding increase in the energy density of matter or radiation
\ Recent observations strongly favour a significant and a positive value of
$\Lambda$ with magnitude $\Lambda(G\hbar/c^{3})\approx10^{-123}$. These
observations suggest on accelerating expansion of the universe, $q<0$.

Therefore, it is considered $G$ and $\Lambda$ as coupling scalars within the
Einstein equations, $R_{ij}-\frac{1}{2}g_{ij}=GT_{ij}-\Lambda g_{ij},$ \ while
the other symbols have their usual meaning and hence the principle of
equivalence then demands that only $g_{ij}$ and not $G$ and $\Lambda$ must
enter the equation of motion of particles and photons. in this way the usual
conservation law, $divT=0,$ holds. Taking the divergence of the Einstein
equations and using the Bianchi identities we obtain the an equation that
controls the variation of $G$ and $\Lambda.$ These are the modified field
equations that allow to take into account a variable $G$ and $\Lambda.$
Nevertheless this approach has some drawbacks, for example, it cannot derived
from a Hamiltonian, although there are several advantages in the approach.

All these works were carry out in the framework of flat
Friedmann--Robertson--Walker (FRW) symmetries. At the same time, all these
works have been extended to more complicated geometries, like for example LRS
Bianchi I as well as Bianchi I models, which represent the simplest
generalization of the flat FRW models (see for example \cite{Bes1}%
-\cite{tony1} in the context of perfect fluids, \cite{Arbab}-\cite{Saha1} in
the context of viscous fluids and \cite{prad1}-\cite{prad2} in the framework
of magnetized viscous fluids). Bianchi I models are important in the study of anisotropies.

Recently , the cosmological implications of a variable speed of light during
the early evolution of the Universe have been considered. Varying speed of
light (VSL) models proposed by Moffat (\cite{moff}) and Albrecht and Magueijo
(\cite{Magueijo0}), in which light was travelling faster in the early periods
of the existence of the Universe, might solve the same problems as inflation.
Hence they could become a valuable alternative explanation of the dynamics and
evolution of our Universe and provide an explanation for the problem of the
variation of the physical \textquotedblleft constants\textquotedblright.
Einstein's field equations (EFE) for FRW spacetime in the VSL theory have been
solved by Barrow (\cite{Barrow-c} and \cite{barrow b1}\ for anisotropic
models), who also obtained the rate of variation of the speed of light
required to solve the flatness and cosmological constant problems (see J.
Magueijo (\cite{Magueijo1}) for a review of these theories).

This model is formulated under the strong assumption that a $c$ variable
(where $c$ stands for the speed of light) does not introduce any corrections
into the curvature tensor, furthermore, such formulation does not verify the
covariance and the Lorentz invariance as well as the resulting field equations
do not verify the Bianchi identities either (see Bassett et al \cite{Bassett}).

Nevertheless, some authors (T. Harko and M. K. Mak \cite{Harko1}, P.P. Avelino
and C.J.A.P. Martins \cite{Avelino} and H. Shojaie et al \cite{Iranies}) have
proposed a new generalization of General Relativity which also allows
arbitrary changes in the speed of light, $c$, and the gravitational constant,
$G$, but in such a way that variations in the speed of light introduces
corrections to the curvature tensor in the Einstein equations in the
cosmological frame. This new formulation is both covariant and Lorentz
invariant and as we will show the resulting field equations (FE) verify the
Bianchi identities. As we have shown in \cite{TonyCarames} this formulation
allows us to obtain the energy conservation equation from the field equations
as in the standard case. Following these ideas we have studied a LRS Bianchi I
with time varying constants in the framework of a viscous fluid as well as for
a perfect fluid (see \cite{tony2}). \ In this works we arrive to the
conclusion that in the early universe, viscous era, from thermodynamical
restrictions, constants $G$ and $c$ are growing time functions while the
cosmological constant is a decreasing time function whose sing depends on the
equation of state. Nevertheless, when we studied the perfect fluid era, we are
not able to determine the behavior of these functions arriving to the
conclusions that $G$ and $c$ may be growing as well as decreasing time
functions while $\Lambda$ is always a decreasing time function.

In a recent paper (see \cite{tony1}) we have developed and compared some well
known tactics (approaches) in order to study and to find exact solutions for a
perfect fluid Bianchi I models with variable $G$ and $\Lambda,$ but trying to
make the lowest number of assumptions or neither. We tried to show that with
these approaches all the usual simplifying hypotheses may be deduced from a
correct mathematical principle and how the useful are each tactic, i.e. to
show the advantages and disadvantages of each approach. We have started
studying this class of models because, as we have mentioned above, there are
many well known exact solutions so we will be able to compare the useful of
our approaches.

The purpose of this paper is to generalize the perfect fluid LRS Bianchi I
model with time varying constants (see \cite{tony2}) taking into account the
effects of a $c$ variable into the field equations. Hence in this paper we are
going to study a Bianchi I model with variable $G,c$ and $\Lambda$ through the
Lie group method (LM), studying the symmetries of the resulting ODE, and
through the self-similarity (SS), matter collineations (MC) as well as
kinematical self-similarity (KSS) hypothesis. We would like to emphasize that
in this work we are more interesting in mathematical respects (as the
integration conditions) than in studying physical consequences. Nevertheless
we consider some observational data in order to rule out some the obtained solutions.

Therefore the paper is divided in the following sections: In section two we
outline the main ingredients of the model as well as the field equations (FE),
under the condition $divT=0.$ Since in this paper we would like to compare the
possible effects of a $c-$var \ with the \textquotedblleft
traditional\textquotedblright\ formulation (i.e. which where such effects are
not taking into account) we need to outline both FE. We will show that such
effect is minimum but exists, although it does not affect to the obtained
solutions through the Lie method. \ In order to apply the (LM) we need to
deduce an ODE. For this purpose we have followed the model proposed by
Kalligas, Wesson and Everitt, \cite{We}, but taking into account some little
differences (obviously here $c=c(t),$ i.e. it is a time varying function). In
this way we have deduced three ODEs. The first one, of second order, which
will be studied in section (\ref{Lie}). This is maybe the main difference, in
this approach, with our previous paper \cite{tony1} where we studied a third
order ODE, but in this occasion as we will show in appendix \ref{appA} there
is no difference between both approaches. In this appendix we will study
through the LM our second ODE while in appendix \ref{appB} we will study
through the LM our third ODE which is a third order ODE but it has been
obtained without considering the effect of $c-$var into the curvature tensor.
In this way we will be able to compare both approaches. As we will see there
is no difference (at least in order of magnitude) between both approaches.

In section three we calculate all the curvature tensors, i.e. Weyl etc... as
well as their invariants, i.e. Kretschmann scalars etc...., but taking into
account the effects of a $c-$var in all the curvature tensors, in our previous
paper (\cite{tony1}) we calculated the same ingredients but in the traditional
way i.e. where $c$ is a true constant. Once we have outlined the FEs then we
go next to study the resulting FEs through the LM, as well as under the SS, MC
and KSS hypotheses.

In section four it is studied through the Lie group tactic the second order
differential equation with four unknowns. \ We seek \ the possible forms that
may take $G,c$ and $\Lambda$ in other to make integrable the ODE. In this way
we find that there are three possibilities, but the question here, is that all
the studied solutions depends of many integrating constants so it is quite
difficult to get information about the real behavior the quantities.
Furthermore, when one try to solve the algebraic system of equations in order
to find the numerical value the exponents of the scale factors finds that the
only possible solution is the flat FRW one (arriving to the same solutions as
the obtained ones in \cite{TonyCarames}), so we arrive to the same scenario as
in our previous paper \cite{tony1}. We think that the followed tactic is too
restrictive, for this reason we are only able to obtain this class of
solutions. Nevertheless there are other Lie approaches as the followed by M.
Szydlowski et al (see \cite{Marek}) which we think that may be more useful
than the followed one here. Trying to improve the obtained solutions, in
appendix \ref{appA}, we will study through the LM the third order ODE, but as
we will show, we arrive to the same solutions and therefore to the same
conclusions. As we have mentioned above, in appendix \ref{appB} we will study
a third order ODE which has been obtained without the assumption of $c-$var
affecting to the curvature tensor. We arrive to the same solutions as the
obtained ones in appendix \ref{appA}, and therefore we conclude that at least
in order of magnitude there is no difference between both approaches.

In section five, we study the model under the self-similarity hypothesis. In
this case, the obtained solution is similar (in order of magnitude) to the
obtained one in the above section (LM with scaling symmetries). Since in this
case, all the obtained solutions, for each quantity, depend on $\left(  \int
c(t)dt\right)  ,$ it is difficult to determine the behavior of each quantity.
Nevertheless, we are able to arrive to some conclusions taking into account
some observational data as for example the hypothesis $q<0,$ (where $q$ stands
for the deceleration parameter) and $\Lambda>0$. Under these considerations we
find that $c$ is a growing time functions while $\Lambda$ is a decreasing time
function whose sing depends on the equation of state $\omega$. With regard to
$G$ we find that it depends on two parameters, the equation of state and $\int
c(t)dt,$ so it may be a decreasing time function as well as a growing time
function depending on the value of these two parameters. In the same way as in
\cite{tony1} we conclude that the exponents of the scale factor must satisfy
the conditions $\sum_{i=1}^{3}\alpha_{i}=1$ and $\sum_{i=1}^{3}\alpha_{i}%
^{2}<1,$ $\forall\omega,$ i.e. for all equation of state, relaxing in this way
the Kasner conditions. Since the model is SS, then, we study the model
studying the matter collineations (MC). In this occasion we only check that
the homothetic vector field verifies the reformulated MC equations (see
\cite{tony1} for details) in order to get information on the behavior of $G,c$
and $\Lambda$, arriving to the same conclusions as in the SS section.

In the last section we reproduce the same tactic, but this time, under the KSS
hypothesis. In this occasion we get a non-singular solution and with the same
behavior for the main quantities as the obtained ones in the above sections.
Before ending this section, we discuss the Kasner like solutions within this
framework arriving to the conclusion that this class of solutions bring us to
get vanishing quantities as well as of having a pathological curvature
behavior since the gravitational entropy tends to infinite. We end with a
brief conclusions.

\section{The Model(s).\label{models}}

Throughout the paper $M$ will denote the usual smooth (connected, Hausdorff,
4-dimensional) spacetime manifold with smooth Lorentz metric $g$ of signature
$(-,+,+,+)$. Thus $M$ is paracompact. A comma, semi-colon and the symbol
$\mathcal{L}$ denote the usual partial, covariant and Lie derivative,
respectively, the covariant derivative being with respect to the Levi-Civita
connection on $M$ derived from $g$. The associated Ricci and stress-energy
tensors will be denoted in component form by $R_{ij}(\equiv R^{c}{}_{jcd})$
and $T_{ij}$ respectively. A diagonal Bianchi I space-time is a spatially
homogeneous space-time which admits an abelian group of isometries $G_{3}$,
acting on spacelike hypersurfaces, generated by the spacelike KVs,
$\mathbf{\xi}_{1}=\partial_{x},\mathbf{\xi}_{2}=\partial_{y},\mathbf{\xi}%
_{3}=\partial_{z}$. In synchronous co-ordinates the metric is:%
\begin{equation}
ds^{2}=-dt^{2}+A_{\mu}^{2}(t)(dx^{\mu})^{2} \label{sx1.2}%
\end{equation}
where the metric functions $A_{1}(t),A_{2}(t),A_{3}(t)$ are functions of the
time co-ordinate only (Greek indices take the space values $1,2,3$ and Latin
indices the space-time values $0,1,2,3$). In this paper we are interested only
in \emph{proper diagonal} Bianchi I space-times (which in the following will
be referred for convenience simply as Bianchi I\ space-times), hence all
metric functions are assumed to be different and the dimension of the group of
isometries acting on the spacelike hypersurfaces is three. Therefore we
consider the Bianchi type I metric as
\begin{equation}
ds^{2}=-c(t)^{2}dt^{2}+X^{2}(t)dx^{2}+Y^{2}(t)dy^{2}+Z^{2}(t)dz^{2},
\label{eq1}%
\end{equation}
see for example (\cite{em}-\cite{Raycha}).

For a perfect fluid with energy-momentum tensor:
\begin{equation}
T_{ij}=\left(  \rho+p\right)  u_{i}u_{j}+pg_{ij}, \label{eq0}%
\end{equation}
where we are assuming an equation of state $p=\omega\rho,\left(
\omega=const.\right)  $. Note that here we have preferred to assume this
equation of state but as we will show in the following sections this equation
may be deduced from the symmetries principles as for example the self-similar
one. The $4-$velocity is defined as follows%
\begin{equation}
u=\left(  \frac{1}{c(t)},0,0,0\right)  .
\end{equation}

The time derivatives of $G,c$ and $\Lambda$ are related by the Bianchi
identities
\begin{equation}
\left(  R_{ij}-\frac{1}{2}Rg_{ij}\right)  ^{;j}=\left(  \frac{8\pi G}{c^{4}%
}T_{ij}-\Lambda g_{ij}\right)  ^{;j}, \label{eq8}%
\end{equation}
in our case we obtain
\begin{equation}
\dot{\rho}+\rho\left(  1+\omega\right)  \left(  \frac{\dot{X}}{X}+\frac
{\dot{Y}}{Y}+\frac{\dot{Z}}{Z}\right)  +\frac{\dot{\Lambda}c^{4}}{8\pi G}%
+\rho\left(  \frac{\dot{G}}{G}-4\frac{\dot{c}}{c}\right)  =0,
\end{equation}
but if we take into account $\ $the condition $\left(  T_{ij}^{;j}=0\right)
,$ it is obtained the following set of equations:
\begin{equation}
\left(  T_{ij}^{;j}=0\right)  \Longleftrightarrow\dot{\rho}+\rho\left(
1+\omega\right)  \left(  \frac{\dot{X}}{X}+\frac{\dot{Y}}{Y}+\frac{\dot{Z}}%
{Z}\right)  =0,
\end{equation}%
\begin{equation}
\frac{\dot{\Lambda}c^{4}}{8\pi G\rho}+\frac{\dot{G}}{G}-4\frac{\dot{c}}{c}=0.
\label{eq9}%
\end{equation}

Therefore the resulting field equations (FE) yield:
\begin{align}
\frac{\dot{X}}{X}\frac{\dot{Y}}{Y}+\frac{\dot{X}}{X}\frac{\dot{Z}}{Z}%
+\frac{\dot{Z}}{Z}\frac{\dot{Y}}{Y}  &  =\frac{8\pi G}{c^{2}}\rho+\Lambda
c^{2},\label{teq_1}\\
\frac{\ddot{Y}}{Y}+\frac{\ddot{Z}}{Z}-\left(  \frac{\dot{Z}}{Z}+\frac{\dot{Y}%
}{Y}\right)  \frac{\dot{c}}{c}+\frac{\dot{Z}}{Z}\frac{\dot{Y}}{Y}  &
=-\frac{8\pi G}{c^{2}}\omega\rho+\Lambda c^{2},\label{teq_2}\\
\frac{\ddot{X}}{X}+\frac{\ddot{Z}}{Z}-\left(  \frac{\dot{X}}{X}+\frac{\dot{Z}%
}{Z}\right)  \frac{\dot{c}}{c}+\frac{\dot{X}}{X}\frac{\dot{Z}}{Z}  &
=-\frac{8\pi G}{c^{2}}\omega\rho+\Lambda c^{2},\label{teq_3}\\
\frac{\ddot{X}}{X}+\frac{\ddot{Y}}{Y}-\left(  \frac{\dot{X}}{X}+\frac{\dot{Y}%
}{Y}\right)  \frac{\dot{c}}{c}+\frac{\dot{X}}{X}\frac{\dot{Y}}{Y}  &
=-\frac{8\pi G}{c^{2}}\omega\rho+\Lambda c^{2},\label{teq_4}\\
\dot{\rho}+\rho\left(  1+\omega\right)  \left(  \frac{\dot{X}}{X}+\frac
{\dot{Y}}{Y}+\frac{\dot{Z}}{Z}\right)   &  =0,\label{teq_5}\\
\frac{\dot{\Lambda}c^{4}}{8\pi G\rho}+\frac{\dot{G}}{G}-4\frac{\dot{c}}{c}  &
=0. \label{teq_6}%
\end{align}

If in eqs. (\ref{teq_2}-\ref{teq_4}) we make $\frac{\dot{c}}{c}=0,$ i.e. we do
not consider the effects of $c-$var into the curvature tensor, then we obtain
the usual FE, see for example \cite{tony1}.

Defining
\begin{equation}
H=\left(  \frac{\dot{X}}{X}+\frac{\dot{Y}}{Y}+\frac{\dot{Z}}{Z}\right)
=3\frac{\dot{R}}{R}\text{ \ and \ \ }R^{3}=XYZ, \label{defH}%
\end{equation}
eq. (\ref{teq_5}) takes the usual form for the conservation equation i.e.
\begin{equation}
\dot{\rho}+\rho\left(  1+\omega\right)  H=0.
\end{equation}

The expansion $\theta$ is defined as follows:
\begin{equation}
\theta:=u_{;i}^{i},\text{ \ \ \ \ \ \ \ \ }\theta=\frac{1}{c(t)}\left(
\frac{\dot{X}}{X}+\frac{\dot{Y}}{Y}+\frac{\dot{Z}}{Z}\right)  =\frac{1}%
{c(t)}H,
\end{equation}
and therefore the acceleration is:%
\begin{equation}
a_{i}=u_{i;j}u^{j},
\end{equation}
in this case $a=0,$ while the shear is defined as follows:%
\begin{equation}
\sigma_{ij}=\frac{1}{2}\left(  u_{i;j}+u_{j;i}+a_{i}u_{j}+a_{j}u_{i}\right)
-\frac{1}{3}\theta h_{ij},
\end{equation}
where $h_{ij}=g_{ij}+u_{i}u_{j}$ is the projection tensor, so%
\begin{equation}
\sigma^{2}=\frac{1}{2}\sigma_{ij}\sigma^{ij},\text{\ \ \ \ \ \ }\sigma
^{2}=\frac{1}{3c^{2}}\left(  \left(  \frac{\dot{X}}{X}\right)  ^{2}+\left(
\frac{\dot{Y}}{Y}\right)  ^{2}+\left(  \frac{\dot{Z}}{Z}\right)  ^{2}%
-\frac{\dot{X}}{X}\frac{\dot{Y}}{Y}-\frac{\dot{X}}{X}\frac{\dot{Z}}{Z}%
-\frac{\dot{Y}}{Y}\frac{\dot{Z}}{Z}\right)  . \label{defshear}%
\end{equation}

\subsection{The main equations.}

In this section we would like to obtain an ODE which allows us to study the
field equations through the Lie method. For this purpose we are following
closely the paper by Kalligas et al (see \cite{We}) and the same steeps
followed in \cite{tony1}.

From eqs. (\ref{teq_2}-\ref{teq_4}) and taking into account eq. (\ref{teq_1}),
it is obtained the following one:
\begin{equation}
\frac{\ddot{X}}{X}+\frac{\ddot{Y}}{Y}+\frac{\ddot{Z}}{Z}-\frac{c^{\prime}}%
{c}\left(  \frac{\dot{X}}{X}+\frac{\dot{Y}}{Y}+\frac{\dot{Z}}{Z}\right)
=-4\pi\left(  1+3\omega\right)  \frac{G}{c^{2}}\rho+\Lambda c^{2}, \label{w1}%
\end{equation}

Now, taking into account eq. (\ref{teq_5}), squaring it and using (\ref{w1})
we get%
\begin{equation}
\left(  \frac{\dot{\rho}}{\rho}\right)  ^{2}=\left(  1+\omega\right)
^{2}\left(  \left(  \frac{\dot{X}}{X}\right)  ^{2}+\left(  \frac{\dot{Y}}%
{Y}\right)  ^{2}+\left(  \frac{\dot{Z}}{Z}\right)  ^{2}+\frac{16\pi}{c^{2}%
}G\rho+2\Lambda c^{2}\right)  , \label{w2}%
\end{equation}
since%
\begin{equation}
H^{2}=\left(  \left(  \frac{\dot{X}}{X}\right)  ^{2}+\left(  \frac{\dot{Y}}%
{Y}\right)  ^{2}+\left(  \frac{\dot{Z}}{Z}\right)  ^{2}+2\left(  \frac{\dot
{X}}{X}\frac{\dot{Y}}{Y}+\frac{\dot{X}}{X}\frac{\dot{Z}}{Z}+\frac{\dot{Z}}%
{Z}\frac{\dot{Y}}{Y}\right)  \right)  , \label{eq-H}%
\end{equation}

The time derivative $\frac{\dot{\rho}}{\rho}$ from eq. (\ref{teq_5}) can now
be expressed in terms of $G,c,\Lambda$ and $\rho$ only by using eqs.
(\ref{w1}) and (\ref{w2}), a straightforward calculation brings us to get the
following expression and hence we get the following expression
\begin{equation}
\ddot{\rho}=\left(  \frac{2+\omega}{1+\omega}\right)  \frac{\dot{\rho}^{2}%
}{\rho}+12\pi\left(  \omega^{2}-1\right)  \frac{G\rho^{2}}{c^{2}}-3\left(
1+\omega\right)  \Lambda c^{2}\rho+\frac{\dot{c}}{c}\dot{\rho}, \label{dorota}%
\end{equation}

\begin{equation}
\ddot{\rho}=K_{1}\frac{\dot{\rho}^{2}}{\rho}+K_{2}\frac{G\rho^{2}}{c^{2}%
}-K_{3}\Lambda c^{2}\rho+\frac{\dot{c}}{c}\dot{\rho}, \label{dorota1}%
\end{equation}
where%
\begin{equation}
K_{1}=\frac{2+\omega}{1+\omega},\qquad K_{2}=12\pi\left(  \omega^{2}-1\right)
,\qquad K_{3}=3\left(  1+\omega\right)  , \label{dorota2}%
\end{equation}
this is the equation that we will study in section (\ref{Lie}) through the Lie
method. As we will see in the appendices (\ref{appA}) and (\ref{appB}) there
is not any advantage if we decide to study the third order ODEs instead of
this one of second order as in our previous paper (see \cite{tony1}), except
that in this case is simpler to study the second order equation instead of the
third order ODE.

Now differentiating eq. (\ref{dorota}) and taking into account eq.
(\ref{teq_6}) we obtain the equation that we will study through the Lie group
method in appendix A. Therefore, we get%
\begin{equation}
\dddot{\rho}=K_{1}\ddot{\rho}\frac{\dot{\rho}}{\rho}-K_{2}\frac{\dot{\rho}%
^{3}}{\rho^{2}}+\frac{G\rho^{2}}{c^{2}}\left[  K_{3}\frac{G^{\prime}}{G}%
-K_{4}\frac{\rho^{\prime}}{\rho}-K_{5}\frac{c^{\prime}}{c}\right]  -K_{6}\rho
cc^{\prime}\Lambda+\dot{\rho}\left(  \frac{c^{\prime\prime}}{c}-\frac
{c^{\prime2}}{c^{2}}\right)  +\frac{c^{\prime}}{c}\left(  \ddot{\rho}%
-\frac{\dot{\rho}^{2}}{\rho}\right)  , \label{defeq}%
\end{equation}
where%
\begin{align}
K_{1}  &  =\frac{5+3\omega}{1+\omega},\qquad K_{2}=\frac{4+2\omega}{1+\omega
},\qquad K_{3}=12\pi\left(  1+\omega\right)  ^{2},\nonumber\\
K_{4}  &  =12\pi\left(  1-\omega^{2}\right)  ,\qquad K_{5}=24\pi\left(
\left(  \omega+2\right)  ^{2}-1\right)  ,\qquad K_{6}=6\left(  1+\omega
\right)  . \label{choren0}%
\end{align}
we are supposing that $\omega\neq-1.$

If we do not consider the effects of $c-$var into the curvature tensor, then
following the same steeps we arrive to the next eq.:
\begin{equation}
\dddot{\rho}=K_{1}\ddot{\rho}\frac{\dot{\rho}}{\rho}-K_{2}\frac{\dot{\rho}%
^{3}}{\rho^{2}}+\frac{G\rho^{2}}{c^{2}}\left[  K_{3}\frac{G^{\prime}}{G}%
-K_{4}\frac{\rho^{\prime}}{\rho}-K_{5}\frac{c^{\prime}}{c}\right]  -K_{6}\rho
cc^{\prime}\Lambda, \label{defeq1}%
\end{equation}
where%
\begin{align}
K_{1}  &  =\frac{5+3\omega}{1+\omega},\qquad K_{2}=\frac{4+2\omega}{1+\omega
},\qquad K_{3}=12\pi\left(  1+\omega\right)  ^{2},\nonumber\\
K_{4}  &  =12\pi\left(  1-\omega^{2}\right)  ,\qquad K_{5}=24\pi\left(
\left(  \omega+2\right)  ^{2}-1\right)  ,\qquad K_{6}=6\left(  1+\omega
\right)  , \label{choren1}%
\end{align}
this eq. will be studied in appendix \ref{appB}. Note that eqs. (\ref{choren1}%
) are the same as eqs. (\ref{choren0}).

As it is observed eq. (\ref{defeq1}) looks simpler than eq. (\ref{defeq}).
Actually, as we will see in appendices (\ref{appA}) and (\ref{appB}) both eqs.
bring us to the same solutions (at least in order of magnitude), so following
this way there is no difference between to study the resulting FE with $c-$var
affecting to the curvature tensor, i.e. eq. (\ref{defeq}), and eq.
(\ref{defeq1}) where we have not take into account such effects.

In this way, it is easy to calculate the shear. Algebra brings us to obtain to
following expression:%
\begin{equation}
\sigma^{2}=\frac{1}{3c^{2}\left(  1+\omega\right)  ^{2}}\left(  \frac
{\dot{\rho}}{\rho}\right)  ^{2}-\left(  8\pi\frac{G}{c^{4}}\rho+\Lambda
\right)  . \label{mshear}%
\end{equation}

\section{Curvature Analysis.\label{curva}}

In this section we calculate some of the curvature invariants (see for example
\cite{Caminati}-\cite{Barrow}) but taking into account the effects of a
$c-$var into the curvature tensors. In (\cite{tony1}) we have calculated the
same invariants but in the traditional way i.e. ignoring the effects of $c(t)$
into the curvature tensors.

The full contraction of the Riemann tensor, i.e. Krestchmann scalars are:%
\begin{equation}
I_{1}:=R_{ijkl}R^{ijkl},
\end{equation}%
\begin{align}
I_{1}  &  =\frac{4}{c^{4}}\left[  \left(  \frac{\ddot{X}}{X}\right)
^{2}-2\frac{\ddot{X}}{X}^{2}\frac{\dot{c}\dot{X}}{cX}+\frac{\dot{c}^{2}\dot
{X}^{2}}{c^{2}X^{2}}+\left(  \frac{\ddot{Y}}{Y}\right)  ^{2}-2\frac{\ddot{Y}%
}{Y}\frac{\dot{Y}}{Y}\frac{\dot{c}}{c}+\left(  \frac{\dot{Y}\dot{c}}%
{Yc}\right)  ^{2}+\left(  \frac{\ddot{Z}}{Z}\right)  ^{2}\right. \nonumber\\
&  \left.  -2\frac{\ddot{Z}}{Z}\frac{\dot{Z}}{Z}\frac{\dot{c}}{c}+\left(
\frac{\dot{Z}}{Z}\frac{\dot{c}}{c}\right)  ^{2}+\left(  \frac{\dot{X}}{X}%
\frac{\dot{Y}}{Y}\right)  ^{2}+\left(  \frac{\dot{X}}{X}\frac{\dot{Z}}%
{Z}\right)  ^{2}+\left(  \frac{\dot{Z}}{Z}\frac{\dot{Y}}{Y}\right)
^{2}\right]  . \label{k1}%
\end{align}%
\begin{equation}
I_{2}:=R_{ij}R^{ij},
\end{equation}%
\begin{align}
I_{2}  &  =\frac{2}{c^{4}}\left[  \left(  \frac{\ddot{X}}{X}\right)
^{2}+\left(  \frac{\ddot{Y}}{Y}\right)  ^{2}+\left(  \frac{\ddot{Z}}%
{Z}\right)  ^{2}+\frac{\ddot{X}}{X}\frac{\ddot{Y}}{Y}+\frac{\ddot{X}}{X}%
\frac{\ddot{Z}}{Z}+\frac{\ddot{Y}}{Y}\frac{\ddot{Z}}{Z}+\frac{\ddot{X}}%
{X}\frac{\dot{X}}{X}\frac{\dot{Y}}{Y}\right. \nonumber\\
&  +\frac{\ddot{X}}{X}\frac{\dot{X}}{X}\frac{\dot{Z}}{Z}+\frac{\ddot{Y}}%
{Y}\frac{\dot{Y}}{Y}\frac{\dot{X}}{X}+\frac{\ddot{Y}}{Y}\frac{\dot{Y}}{Y}%
\frac{\dot{Z}}{Z}+\frac{\ddot{Z}}{Z}\frac{\dot{Z}}{Z}\frac{\dot{X}}{X}%
+\frac{\ddot{Z}}{Z}\frac{\dot{Z}}{Z}\frac{\dot{Y}}{Y}+\left(  \frac{\dot{Z}%
}{Z}\frac{\dot{X}}{X}\right)  ^{2}+\left(  \frac{\dot{Y}}{Y}\frac{\dot{X}}%
{X}\right)  ^{2}\nonumber\\
&  +\left(  \frac{\dot{Y}}{Y}\frac{\dot{Z}}{Z}\right)  ^{2}+\left(  \frac
{\dot{X}}{X}\right)  ^{2}\frac{\dot{Y}}{Y}\frac{\dot{Z}}{Z}+\frac{\dot{X}}%
{X}\left(  \frac{\dot{Y}}{Y}\right)  ^{2}\frac{\dot{Z}}{Z}+\left(  \frac
{\dot{Z}}{Z}\right)  ^{2}\frac{\dot{Y}}{Y}\frac{\dot{X}}{X}-2\frac{\ddot{X}%
}{X}\frac{\dot{X}}{X}\frac{\dot{c}}{c}-\frac{\ddot{X}}{X}\frac{\dot{Y}}%
{Y}\frac{\dot{c}}{c}\nonumber\\
&  -\frac{\ddot{X}}{X}\frac{\dot{Z}}{Z}\frac{\dot{c}}{c}-2\frac{\ddot{Y}}%
{Y}\frac{\dot{Y}}{Y}\frac{\dot{c}}{c}-\frac{\ddot{Z}}{Z}\frac{\dot{X}}{X}%
\frac{\dot{c}}{c}-\frac{\ddot{Y}}{Y}\frac{\dot{X}}{X}\frac{\dot{c}}{c}%
-2\frac{\ddot{Z}}{Z}\frac{\dot{Z}}{Z}\frac{\dot{c}}{c}-\frac{\ddot{Y}}{Y}%
\frac{\dot{Z}}{Z}\frac{\dot{c}}{c}-\frac{\ddot{Z}}{Z}\frac{\dot{Y}}{Y}%
\frac{\dot{c}}{c}\nonumber\\
&  +\left(  \frac{\dot{X}}{X}\frac{\dot{c}}{c}\right)  ^{2}+\left(  \frac
{\dot{Y}}{Y}\frac{\dot{c}}{c}\right)  ^{2}+\left(  \frac{\dot{Z}}{Z}\frac
{\dot{c}}{c}\right)  ^{2}-\left(  \frac{\dot{X}}{X}\right)  ^{2}\frac{\dot{c}%
}{c}\left(  \frac{\dot{Y}}{Y}+\frac{\dot{Z}}{Z}\right)  -\left(  \frac{\dot
{Y}}{Y}\right)  ^{2}\frac{\dot{c}}{c}\left(  \frac{\dot{X}}{X}+\frac{\dot{Z}%
}{Z}\right) \nonumber\\
&  \left.  -\left(  \frac{\dot{Z}}{Z}\right)  ^{2}\frac{\dot{c}}{c}\left(
\frac{\dot{X}}{X}+\frac{\dot{Y}}{Y}\right)  +\left(  \frac{\dot{c}}{c}\right)
^{2}\left(  \frac{\dot{X}}{X}\frac{\dot{Y}}{Y}+\frac{\dot{X}}{X}\frac{\dot{Z}%
}{Z}+\frac{\dot{Y}}{Y}\frac{\dot{Z}}{Z}\right)  \right]  , \label{K2}%
\end{align}
and%
\begin{equation}
R:=R_{i}^{i}=\frac{2}{c^{2}}\left(  \frac{X^{\prime\prime}}{X}+\frac
{Y^{\prime\prime}}{Y}+\frac{Z^{\prime\prime}}{Z}+\frac{X^{\prime}Y^{\prime}%
}{XY}+\frac{Y^{\prime}Z^{\prime}}{YZ}+\frac{X^{\prime}Z^{\prime}}{XZ}\right)
.
\end{equation}

The non-zero components of the Weyl tensor are:%
\begin{align}
C_{1212}  &  =\frac{X^{2}}{6}\left(  -2\frac{X^{\prime\prime}}{X}%
+\frac{X^{\prime}}{X}\left(  2\frac{c^{\prime}}{c}+\frac{Y^{\prime}}{Y}%
+\frac{Z^{\prime}}{Z}\right)  +\left(  \frac{Y^{\prime\prime}}{Y}%
+\frac{Z^{\prime\prime}}{Z}\right)  -\frac{Y^{\prime}}{Y}\left(
\frac{c^{\prime}}{c}+2\frac{Z^{\prime}}{Z}\right)  -\frac{c^{\prime}}{c}%
\frac{Z^{\prime}}{Z}\right)  ,\\
C_{1313}  &  =\frac{Y^{2}}{6}\left(  -2\frac{Y^{\prime\prime}}{Y}%
+\frac{Y^{\prime}}{Y}\left(  2\frac{c^{\prime}}{c}+\frac{X^{\prime}}{X}%
+\frac{Z^{\prime}}{Z}\right)  +\left(  \frac{X^{\prime\prime}}{X}%
+\frac{Z^{\prime\prime}}{Z}\right)  -\frac{X^{\prime}}{X}\left(
\frac{c^{\prime}}{c}+2\frac{Z^{\prime}}{Z}\right)  -\frac{c^{\prime}}{c}%
\frac{Z^{\prime}}{Z}\right)  ,\\
C_{1414}  &  =\frac{Z}{6}\left(  -2\frac{Z^{\prime\prime}}{Z}+\frac{Z^{\prime
}}{Z}\left(  2\frac{c^{\prime}}{c}+\frac{X^{\prime}}{X}+\frac{Y^{\prime}}%
{Y}\right)  +\left(  \frac{X^{\prime\prime}}{X}+\frac{Y^{\prime\prime}}%
{Y}\right)  -\frac{X^{\prime}}{X}\left(  \frac{c^{\prime}}{c}+2\frac
{Y^{\prime}}{Y}\right)  -\frac{c^{\prime}}{c}\frac{Y^{\prime}}{Y}\right)  ,\\
C_{2323}  &  =\frac{X^{2}Y^{2}}{6c^{2}}\left(  2\left(  \frac{Z^{\prime\prime
}}{Z}-\frac{c^{\prime}}{c}\frac{Z^{\prime}}{Z}\right)  -\left(  \frac
{X^{\prime}}{X}+\frac{Y^{\prime}}{Y}\right)  \left(  \frac{Z^{\prime}}%
{Z}-\frac{c^{\prime}}{c}\right)  -\frac{X^{\prime\prime}}{X}-\frac
{Y^{\prime\prime}}{Y}+2\frac{Y^{\prime}}{Y}\frac{X^{\prime}}{X}\right)  ,\\
C_{2424}  &  =\frac{X^{2}Z^{2}}{6c^{2}}\left(  2\left(  \frac{Y^{\prime\prime
}}{Y}-\frac{c^{\prime}}{c}\frac{Y^{\prime}}{Y}\right)  -\left(  \frac
{X^{\prime}}{X}+\frac{Z^{\prime}}{Z}\right)  \left(  \frac{Y^{\prime}}%
{Y}-\frac{c^{\prime}}{c}\right)  -\frac{X^{\prime\prime}}{X}-\frac
{Z^{\prime\prime}}{Z}+2\frac{Z^{\prime}}{Z}\frac{X^{\prime}}{X}\right)  ,\\
C_{3434}  &  =\frac{Y^{2}Z^{2}}{6c^{2}}\left(  2\left(  \frac{X^{\prime\prime
}}{X}-\frac{c^{\prime}}{c}\frac{X^{\prime}}{X}\right)  -\left(  \frac
{Y^{\prime}}{Y}+\frac{Z^{\prime}}{Z}\right)  \left(  \frac{X^{\prime}}%
{X}-\frac{c^{\prime}}{c}\right)  -\frac{Y^{\prime\prime}}{Y}-\frac
{Z^{\prime\prime}}{Z}+2\frac{Y^{\prime}}{Y}\frac{Z^{\prime}}{Z}\right)  ,
\end{align}
where $X^{\prime}:=\dot{X}.$

The non-zero components of the electric part of the Weyl tensor are:%
\begin{align}
E_{22}  &  =\frac{X^{2}}{6c^{2}}\left(  -2\frac{X^{\prime\prime}}{X}%
+\frac{Y^{\prime\prime}}{Y}+\frac{Z^{\prime\prime}}{Z}+\frac{X^{\prime}}%
{X}\left(  \frac{Y^{\prime}}{Y}+\frac{Z^{\prime}}{Z}\right)  -2\frac
{Y^{\prime}}{Y}\frac{Z^{\prime}}{Z}-\frac{c^{\prime}}{c}\left(  2\frac
{X^{\prime}}{X}+\frac{Y^{\prime}}{Y}+\frac{Z^{\prime}}{Z}\right)  \right)  ,\\
E_{33}  &  =\frac{Y^{2}}{6c^{2}}\left(  -2\frac{Y^{\prime\prime}}{Y}%
+\frac{X^{\prime\prime}}{X}+\frac{Z^{\prime\prime}}{Z}+\frac{Y^{\prime}}%
{Y}\left(  \frac{X^{\prime}}{X}+\frac{Z^{\prime}}{Z}\right)  -2\frac
{X^{\prime}}{X}\frac{Z^{\prime}}{Z}-\frac{c^{\prime}}{c}\left(  2\frac
{Y^{\prime}}{Y}+\frac{X^{\prime}}{X}+\frac{Z^{\prime}}{Z}\right)  \right)  ,\\
E_{44}  &  =\frac{Z^{2}}{6c^{2}}\left(  -2\frac{Z^{\prime\prime}}{Z}%
+\frac{X^{\prime\prime}}{X}+\frac{Y^{\prime\prime}}{Y}+\frac{Z^{\prime}}%
{Z}\left(  \frac{X^{\prime}}{X}+\frac{Y^{\prime}}{Y}\right)  -2\frac
{X^{\prime}}{X}\frac{Y^{\prime}}{Y}-\frac{c^{\prime}}{c}\left(  2\frac
{Z^{\prime}}{Z}+\frac{Y^{\prime}}{Y}+\frac{X^{\prime}}{X}\right)  \right)  .
\end{align}

The magnetic part of the Weyl tensor vanish\textbf{\ }%
\begin{equation}
H_{ij}=0.
\end{equation}

The Weyl scalar is%
\begin{equation}
I_{3}=C^{abcd}C_{abcd},
\end{equation}%
\begin{align}
I_{3}  &  =\frac{4}{3c^{2}}\left[  \left(  \frac{X^{\prime\prime}}{X}\right)
^{2}-2\frac{X^{\prime\prime}}{X}\frac{X^{\prime}}{X}\frac{c^{\prime}}%
{c}-2\frac{Y^{\prime\prime}}{Y}\frac{Y^{\prime}}{Y}\frac{c^{\prime}}{c}%
-2\frac{Z^{\prime\prime}}{Z}\frac{Z^{\prime}}{Z}\frac{c^{\prime}}{c}%
+2\frac{Y^{\prime\prime}}{Y}\frac{X^{\prime}}{X}\frac{Z^{\prime}}{Z}%
+2\frac{Z^{\prime\prime}}{Z}\frac{X^{\prime}}{X}\frac{Y^{\prime}}{Y}\right.
\nonumber\\
&  +2\frac{X^{\prime\prime}}{X}\frac{Y^{\prime}}{Y}\frac{Z^{\prime}}{Z}%
-\frac{Z^{\prime\prime}}{Z}\frac{Z^{\prime}}{Z}\left(  \frac{X^{\prime}}%
{X}+\frac{Y^{\prime}}{Y}\right)  +\left(  \frac{c^{\prime}}{c}\right)
^{2}\left(  \left(  \frac{X^{\prime}}{X}\right)  ^{2}+\left(  \frac{Y^{\prime
}}{Y}\right)  ^{2}+\left(  \frac{Z^{\prime}}{Z}\right)  ^{2}\right)
+\nonumber\\
&  \left(  \frac{X^{\prime}}{X}\frac{Z^{\prime}}{Z}\right)  ^{2}+\left(
\frac{X^{\prime}}{X}\frac{Y^{\prime}}{Y}\right)  ^{2}+\left(  \frac{Y^{\prime
}}{Y}\frac{Z^{\prime}}{Z}\right)  ^{2}+\left(  \frac{Y^{\prime\prime}}%
{Y}\right)  ^{2}+\left(  \frac{Z^{\prime\prime}}{Z}\right)  ^{2}%
-\frac{X^{\prime\prime}}{X}\frac{Y^{\prime\prime}}{Y}-\frac{X^{\prime\prime}%
}{X}\frac{Z^{\prime\prime}}{Z}-\frac{Y^{\prime\prime}}{Y}\frac{Z^{\prime
\prime}}{Z}\nonumber\\
&  +\frac{X^{\prime\prime}}{X}\frac{c^{\prime}}{c}\left(  \frac{Y^{\prime}}%
{Y}+\frac{Z^{\prime}}{Z}\right)  +\frac{Y^{\prime\prime}}{Y}\frac{c^{\prime}%
}{c}\left(  \frac{X^{\prime}}{X}+\frac{Z^{\prime}}{Z}\right)  +\frac
{Z^{\prime\prime}}{Z}\frac{c^{\prime}}{c}\left(  \frac{X^{\prime}}{X}%
+\frac{Y^{\prime}}{Y}\right)  -6\frac{X^{\prime}}{X}\frac{Y^{\prime}}{Y}%
\frac{Z^{\prime}}{Z}\frac{c^{\prime}}{c}\nonumber\\
&  \left(  \frac{Z^{\prime}}{Z}\right)  ^{2}\left(  \frac{c^{\prime}}%
{c}\left(  \frac{X^{\prime}}{X}+\frac{Y^{\prime}}{Y}\right)  -\frac{X^{\prime
}}{X}\frac{Y^{\prime}}{Y}\right)  +\left(  \frac{X^{\prime}}{X}\right)
^{2}\left(  \frac{c^{\prime}}{c}\left(  \frac{Z^{\prime}}{Z}+\frac{Y^{\prime}%
}{Y}\right)  -\frac{Y^{\prime}}{Y}\frac{Z^{\prime}}{Z}\right)  +\nonumber\\
&  +\left(  \frac{Y^{\prime}}{Y}\right)  ^{2}\left(  \frac{c^{\prime}}%
{c}\left(  \frac{Z^{\prime}}{Z}+\frac{X^{\prime}}{X}\right)  -\frac{X^{\prime
}}{X}\frac{Z^{\prime}}{Z}\right)  -\left(  \frac{c^{\prime}}{c}\right)
^{2}\frac{X^{\prime}}{X}\left(  \frac{Y^{\prime}}{Y}+\frac{Z^{\prime}}%
{Z}\right)  -\left(  \frac{c^{\prime}}{c}\right)  ^{2}\frac{Y^{\prime}}%
{Y}\frac{Z^{\prime}}{Z}\nonumber\\
&  \left.  -\frac{Y^{\prime\prime}}{Y}\frac{Y^{\prime}}{Y}\left(
\frac{X^{\prime}}{X}+\frac{Z^{\prime}}{Z}\right)  -\frac{X^{\prime\prime}}%
{X}\frac{X^{\prime}}{X}\left(  \frac{Y^{\prime}}{Y}+\frac{Z^{\prime}}%
{Z}\right)  \right]  ,
\end{align}
note that%
\begin{equation}
I_{3}=I_{1}-2I_{2}+\frac{1}{3}R^{2},
\end{equation}
this definition is only valid when $n=4.$

The gravitational entropy is defined as follows (see \cite{gron1}%
-\cite{gron2}):%
\begin{equation}
P^{2}=\frac{I_{3}}{I_{2}}=\frac{I_{1}-2I_{2}-\frac{1}{3}R^{2}}{I_{2}}%
=\frac{I_{1}}{I_{2}}+\frac{1}{3}\frac{R^{2}}{I_{2}}-2.
\end{equation}

\section{Lie Method.\label{Lie}}

Therefore we are interesting in studying through the Lie method eq.
(\ref{dorota1}) i.e.%
\begin{equation}
\ddot{\rho}=K_{1}\frac{\dot{\rho}^{2}}{\rho}+K_{2}\frac{G\rho^{2}}{c^{2}%
}-K_{3}\Lambda c^{2}\rho+\frac{\dot{c}}{c}\dot{\rho}, \label{nuria}%
\end{equation}
where $\left(  K_{i}\right)  _{i=1}^{3}\in\mathbb{R},$ are given by eqs.
(\ref{dorota2}), in particular we seek the forms of $G,c$ and $\Lambda,$ for
which our field equations admit symmetries i.e. are integrable (see for
example \cite{Ibra}-\cite{TonyCas}). We would like to stress that eq.
(\ref{nuria}) is very similar to the studied one in the context of FRW
symmetries (see \cite{TonyCarames}).

For this purpose, and following the standard procedure we need to solve the
following system of PDEs:%
\begin{align}
K_{1}\xi_{\rho}+\rho\xi_{\rho\rho}  &  =0,\label{nu1}\\
\eta_{\rho\rho}-2\xi_{t\rho}+\frac{K_{1}}{\rho}\left(  \eta-\rho\eta_{\rho
}\right)  -2\frac{c^{\prime}}{c}\xi_{\rho}  &  =0,\label{nur2}\\
\rho^{2}c^{2}\left[  2\eta_{t\rho}-\xi_{tt}+3\rho\xi_{\rho}\left(  c^{2}%
K_{3}\Lambda-\rho K_{2}\frac{G}{c^{2}}\right)  -2K_{1}\frac{\eta_{t}}{\rho
}+\xi\left(  -\frac{c^{\prime\prime}}{c}+\frac{c^{\prime2}}{c^{2}}\right)
-\xi_{t}\frac{c^{\prime}}{c}\right]   &  =0,\label{nur3}\\
\rho^{2}\left[  \eta_{tt}c^{2}-c\eta_{t}c^{\prime}+K_{2}\rho^{2}G\left(
-\xi\left(  \frac{G^{\prime}}{G}-2\frac{c^{\prime}}{c}\right)  -2\frac{\eta
}{\rho}-2\xi_{t}+\eta_{\rho}\right)  +K_{3}c^{4}\left(  \Lambda\left(
\eta-\rho\eta_{\rho}\right)  +\rho\Lambda\left(  2\xi_{t}+\xi\left(
\frac{\Lambda^{\prime}}{\Lambda}+2\frac{c^{\prime}}{c}\right)  \right)
\right)  \right]   &  =0. \label{nur4}%
\end{align}

Imposing the symmetry $X=\left(  at+e\right)  \partial_{t}+b\rho\partial
_{\rho},$ i.e. $\left(  \xi=at+e,\ \eta=b\rho\right)  ,$ we get the following
restrictions for $G(t),c(t)$ and $\Lambda(t)$. Note that constants $\left(
a,b,e\right)  \in\mathbb{R}$, where $\left[  a\right]  =\left[  b\right]  =1,$
i.e. they are dimensionless while $\left[  e\right]  =T,$ with respect to the
dimensional base $B=\left\{  L,M,T\right\}  ,$ i.e. constant $e$ has
dimensions of time, $T.$

From eq. (\ref{nur3}) we get%
\begin{equation}
\frac{c^{\prime\prime}}{c^{\prime}}-\frac{c^{\prime}}{c}=-\frac{2a}{at+e}.
\label{silv1}%
\end{equation}

Now, from eq. (\ref{nur4}) it is obtained%
\begin{equation}
\frac{G^{\prime}}{G}-2\frac{c^{\prime}}{c}=-\frac{2a+b}{at+e}, \label{silv2}%
\end{equation}
and
\begin{equation}
\frac{\Lambda^{\prime}}{\Lambda}+2\frac{c^{\prime}}{c}=-\frac{2a}{at+e}.
\label{silv3}%
\end{equation}

Therefore, for the different values of the constants $\left(  a,b,e\right)  $
we will be a able to find different behaviors for the functions $G(t),c(t)$
and $\Lambda(t),$ and hence to integrate eq. (\ref{nuria}).

\subsection{Scale symmetry.\label{liescale}}

Making $e=0,$ $i.e.$ considering only $\left(  \xi=at,\ \eta=b\rho\right)  , $
we obtain the scale symmetry, $X=at\partial_{t}+b\rho\partial_{\rho}, $ so
eqs. (\ref{silv1}-\ref{silv3}) yield:
\begin{align}
\frac{c^{\prime\prime}}{c^{\prime}}-\frac{c^{\prime}}{c}  &  =-\frac{1}%
{t},\label{2elena}\\
\frac{G^{\prime}}{G}-2\frac{c^{\prime}}{c}  &  =-\frac{b+2a}{at}%
,\qquad\Longrightarrow\qquad\frac{G}{c^{2}}=Bt^{-\left(  2+\frac{b}{a}\right)
},\\
\frac{\Lambda^{\prime}}{\Lambda}+2\frac{c^{\prime}}{c}  &  =-\frac{2}%
{t},\qquad\Longrightarrow\qquad\Lambda c^{2}=\tilde{B}t^{-2},
\end{align}
where $B,\tilde{B}\in\mathbb{R},$ therefore we get
\begin{align}
c  &  =c_{0}t^{c_{1}},\qquad c_{1},c_{0}\in\mathbb{R},\label{2cscale}\\
G  &  =G_{0}t^{2(c_{1}-1)-\frac{b}{a}},\qquad G_{0}\in\mathbb{R}%
^{+},\label{2Gscale}\\
\Lambda &  =\Lambda_{0}t^{-2(c_{1}+1)},\qquad\Lambda_{0}\in\mathbb{R},
\label{2Lscale}%
\end{align}
where we assume that $G_{0}>0.$

The invariant solution for the energy density is:
\begin{equation}
\frac{bdt}{at}=\frac{d\rho}{\rho}\qquad\Longrightarrow\qquad\rho=\rho
_{0}t^{b/a},
\end{equation}
and for physical reasons we impose the condition, $ab<0$ then $b<0.$ We have
consider the invariant solution since, as we already known, the most general
solution for eq. (\ref{nuria}) with the constrains given by eqs.
(\ref{2cscale}-\ref{2Lscale}) usually is an unphysical solution (i.e. it lacks
of physical meaning, see (\cite{TonyCas})). Furthermore, as we will show in
the next section this kind of spacetime is self-similar which means that all
the quantities follow a power law, as in this case with the invariant
solution, see (\cite{Jantzen}-\cite{Wainwrit}).

If we make that this solution verifies eq. (\ref{nuria}) with $c(t),G(t)$ and
$\Lambda(t)$ given by eqs. (\ref{2cscale}-\ref{2Lscale}), we find the value of
constant $\rho_{0},$ so%
\begin{equation}
\rho_{0}=-\left[  \frac{c_{0}^{2}\left(  b^{2}+ab\left(  1+\omega\right)
(c_{1}+1)-3c_{0}^{2}\Lambda_{0}a^{2}\left(  1+\omega\right)  ^{2}\right)
}{12\pi a^{2}G_{0}\left(  1+\omega\right)  ^{2}\left(  \omega-1\right)
}\right]  , \label{ayla}%
\end{equation}
with the only restriction $\omega\neq-1,1.$ Note that $ab<0,$ so we need to
choice constants $\left(  c_{0},G_{0},\Lambda_{0}\right)  $ such that
$\rho_{0}>0.$

\begin{remark}
As we can see, it is verified the relationship $\frac{G\rho}{c^{2}%
}\thickapprox t^{-2},$ as well as $\Lambda c^{2}\thickapprox t^{-2}.$
\end{remark}

Therefore, at this time we have the following behavior for $G(t)$
\begin{equation}
G(t)=G_{0}t^{2(c_{1}-1-\frac{b}{2a})},\qquad G\thickapprox\left\{
\begin{array}
[c]{l}%
\text{decreasing if }c_{1}<1+b/2a,\\
\text{constant if }c_{1}=1+b/2a,\\
\text{growing if }c_{1}>1+b/2a,
\end{array}
\right.  ,
\end{equation}
while $\Lambda$ behaves as follows:%
\begin{equation}
\Lambda=\Lambda_{0}t^{-2(c_{1}+1)},,\qquad\Lambda\thickapprox\left\{
\begin{array}
[c]{l}%
\text{decreasing if }c_{1}>-1,\\
\text{constant if }c_{1}=-1,\\
\text{growing if }c_{1}<-1,
\end{array}
\right.  ,
\end{equation}
therefore $\left(  c_{1}+1\right)  >0\Longrightarrow c_{1}\in\left(
-1,\infty\right)  .$ But we have not any information about the sign of
$\ \Lambda_{0}. $

With regard to $H=3\frac{R^{\prime}}{R}$ we find that
\begin{equation}
R=R_{0}\rho^{-1/3\left(  1+\omega\right)  }=R_{0}t^{-b/3a\left(
1+\omega\right)  },\qquad\Longrightarrow\qquad XYZ=R_{0}t^{-b/a\left(
1+\omega\right)  },
\end{equation}
and assuming that the functions $\left(  X,Y,Z\right)  $ follow a power law
(i.e. $X=X_{0}t^{\alpha_{1}})$ then we get the following result $Kt^{\alpha
}=R_{0}t^{-b/a\left(  1+\omega\right)  },$ and therefore, $\sum_{i}^{3}%
\alpha_{i}=\alpha=-\left(  b/a\left(  1+\omega\right)  \right)  ,$ where we
may assume that $\left(  \alpha_{i}\right)  >0,\forall i$ and $\left(
\alpha_{i}\neq\alpha_{j}\right)  $ although $\left(  \alpha_{i}\rightarrow
\alpha_{j}\right)  $ when $t\rightarrow\infty,$ with $i\neq j,$ i.e. we expect
that the model isotropize and collapses to a FRW model, but we have not more
information about this behavior.

The shear is calculated as follows, from eq. (\ref{mshear}) we get:%
\begin{equation}
\sigma^{2}=\sigma_{0}^{2}t^{-2(c_{1}+1)},\text{\qquad with\qquad}\sigma
_{0}^{2}=\frac{b^{2}+2ab(c_{1}+1)-3a^{2}\Lambda_{0}c_{0}^{2}\left(
1+\omega\right)  ^{2}}{3a^{2}\left(  \omega^{2}-1\right)  },
\end{equation}
but in this case it is quite difficult to know if $\sigma_{0}^{2}\neq0$ or
$\sigma_{0}^{2}=0.$ Therefore, at this time, we cannot to rule out this
solution as in the case of $G$ and $\Lambda$ time varying (see (\cite{tony1}))
where $\sigma_{0}^{2}=0.$ The only important obtained restriction is $c_{1}%
\in\left(  -1,\infty\right)  .$

\subsection{Exponential behavior.\label{lieexp}}

Making $a=0,$ we have $\left(  \xi=e,\,\eta=b\rho\right)  ,$ so following the
same steeps, we have to integrate eqs. (\ref{silv1}-\ref{silv3}), hence
\begin{align}
\frac{c^{\prime\prime}}{c^{\prime}}-\frac{c^{\prime}}{c}  &
=0,\label{2elena1}\\
\frac{G^{\prime}}{G}-2\frac{c^{\prime}}{c}  &  =-\frac{b}{e},\qquad
\Longrightarrow\qquad\frac{G}{c^{2}}=\exp\left(  -\frac{b}{e}t\right)  ,\\
\frac{\Lambda^{\prime}}{\Lambda}+2\frac{c^{\prime}}{c}  &  =0,\qquad
\Longrightarrow\qquad\Lambda c^{2}=const,
\end{align}
therefore we get
\begin{align}
c  &  =c_{0}\exp(c_{1}t),\qquad c_{0},c_{1}\in\mathbb{R}^{+},\label{2cexp}\\
G  &  =G_{0}\exp\left[  \left(  2c_{1}-\frac{b}{e}\right)  t\right]  ,\qquad
G_{0}\in\mathbb{R}^{+},\label{2Gexp}\\
\Lambda &  =\Lambda_{0}\exp(-2c_{1}t),\qquad\Lambda_{0}\in\mathbb{R},
\label{2Lexp}%
\end{align}
where we assume that $c_{0},G_{0}>0.$ From eq. (\ref{2Lexp}) we find that
$c_{1}>0,$ otherwise $\Lambda$ will be a growing function on time.

The invariant solution for the energy density is:
\begin{equation}
\rho=\rho_{0}\exp(\frac{b}{e}t),
\end{equation}
with the restriction, $eb<0$ with $b<0,$ from physical considerations. In
order to calculate the value of constant $\rho_{0},$ this solution must
verifies eq. (\ref{nuria}) with $c(t),G(t)$ and $\Lambda(t)$ given by eqs.
(\ref{2cexp}-\ref{2Lexp}), finding in this way that constant $\rho_{0}$
yields$,$%
\begin{equation}
\rho_{0}=-\left[  \frac{c_{0}^{2}\left(  b^{2}+eb\left(  1+\omega\right)
c_{1}-3c_{0}^{2}\Lambda_{0}e^{2}\left(  1+\omega\right)  ^{2}\right)  }{12\pi
e^{2}G_{0}\left(  1+\omega\right)  ^{2}\left(  \omega-1\right)  }\right]  .
\label{2ro0exp}%
\end{equation}

With regard to $H$ we find that
\begin{equation}
R=R_{0}\rho^{-1/3\left(  1+\omega\right)  }=R_{0}\exp\left(  -\frac
{b}{3e(\omega+1)}t\right)  ,
\end{equation}

The shear is calculated as follows$.$%
\begin{equation}
\sigma^{2}=\sigma_{0}^{2}\exp(-2c_{1}t),\qquad\text{with \qquad}\sigma_{0}%
^{2}=\frac{b^{2}+2ebc_{1}-3e^{2}\Lambda_{0}c_{0}^{2}\left(  1+\omega\right)
^{2}}{3e^{2}\left(  \omega^{2}-1\right)  }.
\end{equation}

\begin{remark}
As we can see, it is impossible to have any information about the real
behavior of the quantities since they depend on several integration constants.
We suppose that this model is unphysical but we have not any way of rule it out.
\end{remark}

\subsection{Solution with the full symmetry.\label{liefull}}

In this case we have the full symmetry i.e. $\left(  \xi=at+e,\ \eta
=b\rho\right)  ,$ with $\left[  e\right]  =T,$ we have to integrate eqs.
(\ref{silv1}-\ref{silv3}), so
\begin{align}
\frac{c^{\prime\prime}}{c^{\prime}}-\frac{c^{\prime}}{c}  &  =-\frac{a}%
{at+e},\label{2elena2}\\
\frac{G^{\prime}}{G}-2\frac{c^{\prime}}{c}  &  =-\frac{b+2a}{at+e}%
,\qquad\Longrightarrow\qquad\frac{G}{c^{2}}=\left(  at+e\right)  ^{-\left(
\frac{b}{a}+2\right)  },\\
\frac{\Lambda^{\prime}}{\Lambda}+2\frac{c^{\prime}}{c}  &  =-\frac{2a}%
{at+e},\qquad\Longrightarrow\qquad\Lambda c^{2}=\left(  at+e\right)  ^{-2},
\end{align}
therefore we get
\begin{align}
c  &  =c_{0}\left(  at+e\right)  ^{c_{1}/a},\qquad c_{1},c_{0}\in
\mathbb{R},\label{2cfull}\\
G  &  =G_{0}\left(  at+e\right)  ^{2\frac{c_{1}}{a}-2-\frac{b}{a}},\qquad
G_{0}\in\mathbb{R}^{+},\label{2Gfull}\\
\Lambda &  =\Lambda_{0}\left(  at+e\right)  ^{-2(1+c_{1}/a)},\qquad\Lambda
_{0}\in\mathbb{R}, \label{2Lfull}%
\end{align}
where we assume that $c_{0},G_{0}>0.$

The invariant solution for the energy density is%
\begin{equation}
\rho=\rho_{0}(at+e)^{b/a}, \label{2rhofull}%
\end{equation}
where we need to impose the physical constrain such that $ab<0$ then $b<0.$ In
order to find the constant $\rho_{0},$ we make that solution (\ref{2rhofull})
verifies eq. (\ref{nuria}) with $G(t),c(t)$ and $\Lambda(t)$ given by eq.
(\ref{2cfull}-\ref{2Lfull}), finding in this way that the value of the
numerical constant $\rho_{0},$ yields%
\begin{equation}
\rho_{0}=-\left[  \frac{c_{0}^{2}\left(  b^{2}+b\left(  c_{1}+a\right)
\left(  1+\omega\right)  -3c_{0}^{2}\Lambda_{0}\left(  1+\omega\right)
^{2}\right)  }{12\pi G_{0}\left(  1+\omega\right)  ^{2}\left(  \omega
-1\right)  }\right]  ,
\end{equation}
we assume that $\omega\neq-1.$ As it is observed, this is a nonsingular
solution since when $t\rightarrow0$ if $e\neq0,$ then $\rho\neq\infty.$ As in
the above cases, it is verified the condition $G\rho/c^{2}\thickapprox
(at+e)^{-2},$ as well as, $\Lambda c^{2}\thickapprox(at+e)^{-2}.$

Therefore we have the following behavior for $G(t):$
\begin{equation}
G(t)=G_{0}(at+e)^{-(2a+b)/a},\qquad G\thickapprox\left\{
\begin{array}
[c]{l}%
\text{decreasing if }c_{1}<1+b/2a,\\
\text{constant if }c_{1}=1+b/2a,\\
\text{growing if }c_{1}>1+b/2a,
\end{array}
\right.  ,
\end{equation}
note that if $2a=-b,$ then $c_{1}=0.$ with $\left(  a,e>0,b<0\right)  $. As we
can see, it is obtained the same behavior as in the scale symmetry solution.

Lambda behaves as $\Lambda=\Lambda_{0}\left(  at+e\right)  ^{-2(1+c_{1}/a)},$
so $\left(  1+c_{1}/a\right)  >0,$ i.e. $\left\vert c_{1}\right\vert
<\left\vert a\right\vert ,$ which means that $c_{1}\in\left(  -a,\infty
\right)  ,$ finding a bit difference with respect to the scale symmetry solution.

With regard to the quantity $H,$ we find from eq. (\ref{defH}) that
\begin{equation}
R=R_{0}\rho^{-1/3\left(  1+\omega\right)  }=R_{0}(at+e)^{-b/3a\left(
1+\omega\right)  },\qquad\Longrightarrow\qquad XYZ=R_{0}(at+e)^{-b/a\left(
1+\omega\right)  },
\end{equation}
so (following the same argument as above) the functions $\left(  X,Y,Z\right)
$ follow a power law (i.e. $X=X_{0}(at+e)^{\alpha_{1}},$ etc...$)$ it is found
that, $K(at+e)^{\alpha}=R_{0}(at+e)^{-b/a\left(  1+\omega\right)  },$ and
hence, $\sum_{i}^{3}\alpha_{i}=\alpha=-\left(  b/a\left(  1+\omega\right)
\right)  ,$ where we may \textquotedblleft assume\textquotedblright\ that
$\left(  \alpha_{i}\right)  >0,\forall i$ and $\left(  \alpha_{i}\neq
\alpha_{j}\right)  $ although $\left(  \alpha_{i}\rightarrow\alpha_{j}\right)
$ when $t\rightarrow\infty,$ and $i\neq j.$

The shear has the following behavior$.$%
\begin{equation}
\sigma^{2}=\frac{b^{2}+2b\left(  c_{1}+a\right)  -3c_{0}^{2}\Lambda_{0}\left(
1+\omega\right)  ^{2}}{3\left(  \omega^{2}-1\right)  }(at+e)^{-2(1+c_{1}/a)}.
\end{equation}

At this point it seems that we have found a physical solution that depends on
the value of constants $a$ and $b.$ But, how to calculate the value of
constants $\left(  \alpha_{i}\right)  _{i=1}^{3}?.$ Simply all these results
must satisfy the FE, hence
\begin{align}
\alpha_{1}\alpha_{2}+\alpha_{1}\alpha_{3}+\alpha_{2}\alpha_{3}  &  =8\pi
\frac{G_{0}}{c_{0}^{2}}\rho_{0}+\Lambda_{0}c_{0}^{2},\\
\alpha_{2}\left(  \alpha_{2}-1\right)  +\alpha_{3}\left(  \alpha_{3}-1\right)
+\alpha_{3}\alpha_{2}-c_{1}\left(  \alpha_{3}+\alpha_{2}\right)   &
=-8\pi\frac{G_{0}}{c_{0}^{2}}\omega\rho_{0}+\Lambda_{0}c_{0}^{2},\\
\alpha_{1}\left(  \alpha_{1}-1\right)  +\alpha_{3}\left(  \alpha_{3}-1\right)
+\alpha_{3}\alpha_{1}-c_{1}\left(  \alpha_{1}+\alpha_{3}\right)   &
=-8\pi\frac{G_{0}}{c_{0}^{2}}\omega\rho_{0}+\Lambda_{0}c_{0}^{2},\\
\alpha_{2}\left(  \alpha_{2}-1\right)  +\alpha_{1}\left(  \alpha_{1}-1\right)
+\alpha_{1}\alpha_{2}-c_{1}\left(  \alpha_{1}+\alpha_{2}\right)   &
=-8\pi\frac{G_{0}}{c_{0}^{2}}\omega\rho_{0}+\Lambda_{0}c_{0}^{2},
\end{align}
which solution is:%
\begin{equation}
\alpha_{1}=\alpha_{2}=\alpha_{3}=\sqrt{\frac{8\pi G_{0}\rho_{0}}{3c_{0}^{2}%
}+\frac{\Lambda_{0}c_{0}^{2}}{3}},\qquad c_{1}=-1+\frac{4\pi G_{0}\rho
_{0}(1+\omega)}{c_{0}\sqrt{\frac{8\pi G_{0}\rho_{0}}{3}+\frac{\Lambda_{0}%
c_{0}^{4}}{3}}},
\end{equation}
finding that this kind of solutions lack of any interest, it is the flat FRW
one. This solution was obtained by Einstein\&de Sitter (\cite{EdS}) in 1932
for $\omega=0$, and later by Harrison (\cite{Harrison}) $\forall\omega.$

We would like to point out that, at least, this solution is consistent with
the already obtained one in (\cite{TonyCarames}), where we studied a perfect
fluid with time varying constants (as here, i.e. taking into account the
possible effects of a c-var into the curvature tensor) but in the context of
the flat FRW symmetries. We think that the followed methods is too restrictive
and for this reason we are only able to get this class of solutions. As we
have pointed out above eq. (\ref{nuria}) is quite similar to the FRW case
studied in (\cite{TonyCarames}) and for this reason with this approach we are
only able to obtain FRW-like solution.

There are others Lie group approaches, as for example, the developed by M.
Szydlowski et al (see \cite{Marek}), maybe if we follow this approach we would
be able to get other class of solutions as it is expected studying this kind
of spacetimes i.e. to get, for example, a Kasner-like solution.

\section{Self-similar solution.\label{SS}}

In general relativity, the term self-similarity can be used in two ways. One
is for the properties of spacetimes, the other is for the properties of matter
fields. These are not equivalent in general. The self-similarity in general
relativity was defined for the first time by Cahill and Taub (see \cite{CT},
and for general reviews \cite{21}-\cite{Hall}). Self-similarity is defined by
the existence of a homothetic vector ${V}$ in the spacetime, which satisfies
\begin{equation}
\mathcal{L}_{V}g_{ij}=2\alpha g_{ij}, \label{gss1}%
\end{equation}
where $g_{ij}$ is the metric tensor, $\mathcal{L}_{V}$ denotes Lie
differentiation along ${V}$ and $\alpha$ is a constant. This is a special type
of conformal Killing vectors. This self-similarity is called homothety. If
$\alpha\neq0$, then it can be set to be unity by a constant rescaling of ${V}%
$. If $\alpha=0$, i.e. $\mathcal{L}_{V}g_{ij}=0$, then ${V}$ is a Killing vector.

Homothety is a purely geometric property of spacetime so that the physical
quantity does not necessarily exhibit self-similarity such as $\mathcal{L}%
_{V}Z=dZ$, where $d$ is a constant and $Z$ is, for example, the pressure, the
energy density and so on. From equation (\ref{gss1}) it follows that
\begin{equation}
\mathcal{L}_{V}R^{i}\,_{jkl}=0,
\end{equation}
and hence
\begin{equation}
\mathcal{L}_{V}R_{ij}=0,\qquad\mathcal{L}_{V}G_{ij}=0. \label{mattercoll}%
\end{equation}
A vector field ${V}$ that satisfies the above equations is called a curvature
collineation, a Ricci collineation and a matter collineation, respectively. It
is noted that such equations do not necessarily mean that ${V}$ is a
homothetic vector. We consider the Einstein equations
\begin{equation}
G_{ij}=8\pi GT_{ij}, \label{einstein}%
\end{equation}
where $T_{ij}$ is the energy-momentum tensor.

If the spacetime is homothetic, the energy-momentum tensor of the matter
fields must satisfy
\begin{equation}
\mathcal{L}_{V}T_{ij}=0, \label{emcoll}%
\end{equation}
through equations~(\ref{einstein}) and (\ref{mattercoll}). For a perfect fluid
case, the energy-momentum tensor takes the form of eq. (\ref{eq0}) i.e.
$T_{ij}=(p+\rho)u_{i}u_{j}+pg_{ij},$where $p$ and $\rho$ are the pressure and
the energy density, respectively. Then, equations~(\ref{gss1}) and
(\ref{emcoll}) result in
\begin{equation}
\mathcal{L}_{V}u^{i}=-\alpha u^{i},\qquad\mathcal{L}_{V}\rho=-2\alpha
\rho,\qquad\mathcal{L}_{V}p=-2\alpha p. \label{ssmu}%
\end{equation}
As shown above, for a perfect fluid, the self-similarity of the spacetime and
that of the physical quantity coincide. However, this fact does not
necessarily hold for more general matter fields. Thus the self-similar
variables can be determined from dimensional considerations in the case of
homothety. Therefore, we can conclude homothety as the general relativistic
analogue of complete similarity.

From the constraints (\ref{ssmu}), we can show that if we consider the
barotropic equation of state, i.e., $p=f(\rho)$, then the equation of state
must have the form $p=\omega\rho$, where $\omega$ is a constant. This class of
equations of state contains a stiff fluid ($\omega=1$) as special cases,
whiting this theoretical framework. There are many papers devoted to study
Bianchi I models (in different context) assuming the hypothesis of
self-similarity (see for example \cite{HW}-\cite{griego}) but here, we would
like to try to show how taking into account this class of hypothesis one is
able to find exact solutions to the field equations within the framework of
the time varying constants.

The homothetic equations are: $L_{V}g=2g,$ finding that the homothetic vector
field is in this case:%
\begin{equation}
X=\left(  \frac{\int cdt}{c(t)}\right)  \partial_{t}+\left(  1-\frac{\int
cdt}{c(t)}\frac{\dot{X}}{X}\right)  x\partial_{x}+\left(  1-\frac{\int
cdt}{c(t)}\frac{\dot{Y}}{Y}\right)  y\partial_{y}+\left(  1-\frac{\int
cdt}{c(t)}\frac{\dot{Z}}{Z}\right)  z\partial_{z}, \label{HO1}%
\end{equation}
iff the following ODE is satisfied
\begin{equation}
\left(  X\dot{X}c^{2}-X\dot{X}c^{\prime}\int cdt+cX\ddot{X}\int cdt-\left(
\dot{X}\right)  ^{2}c\int cdt\right)  \frac{x}{c^{2}}=0, \label{helen}%
\end{equation}
and so on with respect to $\left(  Y,y\right)  $ and $\left(  Z,z\right)  .$

As it is observed from eq. (\ref{helen}) if we simplify this ODE it is
obtained the following one%
\begin{equation}
\frac{H_{1}^{\prime}}{H_{1}}=\frac{c^{\prime}}{c}-\frac{c}{\int cdt}%
,\Longrightarrow H_{1}=\alpha_{1}\frac{c}{\int cdt},\qquad\Longrightarrow
\qquad X=X_{0}\left(  \int cdt\right)  ^{\alpha_{1}},
\end{equation}
with $\alpha_{1}\in\mathbb{R},$ etc....with regard to the others scale
factors. Note that $^{\prime}:=\frac{d}{dt}:=dot.$ i.e. $X^{\prime}=\dot{X}.$
So, we have%
\begin{equation}
X=X_{0}\left(  \int cdt\right)  ^{\alpha_{1}},\qquad Y=Y_{0}\left(  \int
cdt\right)  ^{\alpha_{2}},\qquad Z=Z_{0}\left(  \int cdt\right)  ^{\alpha_{3}%
}, \label{scalesSS}%
\end{equation}
with $\left(  \alpha_{i}\right)  _{i=1}^{3}\in\mathbb{R},$ note that at this
time we have not any information about the possible values and their signs of
the numerical constants $\left(  \alpha_{i}\right)  _{i=1}^{3}.$

\begin{remark}
Note that if $c=const.$ we regain the usual homothetic vector field. i.e.%
\begin{equation}
X=t\partial_{t}+\left(  1-t\frac{\dot{X}}{X}\right)  x\partial_{x}+\left(
1-t\frac{\dot{Y}}{Y}\right)  y\partial_{y}+\left(  1-t\frac{\dot{Z}}%
{Z}\right)  z\partial_{z}, \label{homoc}%
\end{equation}
while the scale factors behave as%
\begin{equation}
X=X_{0}\left(  t\right)  ^{\alpha_{1}},\qquad Y=Y_{0}\left(  t\right)
^{\alpha_{2}},\qquad Z=Z_{0}\left(  t\right)  ^{\alpha_{3}},
\end{equation}
as in the case with only $G$ and $\Lambda$ variable (see \cite{tony1}).
\end{remark}

Since%
\begin{equation}
H_{i}=\alpha_{i}\frac{c}{\int cdt}\qquad\Longrightarrow\qquad H=\alpha\frac
{c}{\int cdt},\qquad\alpha=\sum_{i=1}^{3}\alpha_{i}, \label{HSS}%
\end{equation}
finding in this way, from eq. (\ref{teq_5}), the behavior of the energy
density i.e.%
\begin{equation}
\rho=\rho_{0}\left(  \int cdt\right)  ^{-(1+\omega)\alpha}. \label{rhoSS}%
\end{equation}

In the same way it is easily calculated the shear%
\begin{equation}
\sigma^{2}=\frac{1}{3c^{2}}\left(  H_{1}^{2}+H_{2}^{2}+H_{3}^{2}-H_{1}%
H_{2}-H_{1}H_{3}-H_{2}H_{3}\right)  =\frac{1}{3}\left(  \sum_{i=1}^{3}%
\alpha_{i}^{2}-\sum_{i\neq j}\alpha_{i}\alpha_{j}\right)  \left(  \int
cdt\right)  ^{-2}.
\end{equation}

As it is observed all the quantities depend on $c(t)$, so only rest to
calculate $G$ and $\Lambda$.

From eqs. (\ref{teq_1}, \ref{HSS} and \ref{rhoSS}) we get:%
\begin{equation}
A\left(  \frac{c}{\int c}\right)  ^{2}=\frac{8\pi G}{c^{2}}\rho_{0}\left(
\int c\right)  ^{-\gamma}+\Lambda c^{2},
\end{equation}
where we have written, for simplicity, $\int c$ instead of $\int cdt,$ and
$A=\left(  \alpha_{1}\alpha_{2}+\alpha_{1}\alpha_{3}+\alpha_{2}\alpha
_{3}\right)  ,\gamma=(1+\omega)\alpha,$ therefore%
\begin{equation}
\Lambda^{\prime}=-\frac{2Ac}{\left(  \int c\right)  ^{3}}-\frac{8\pi\rho_{0}%
G}{c^{4}\left(  \int c\right)  ^{\gamma}}\left[  \frac{G^{\prime}}{G}%
-4\frac{c^{\prime}}{c}-\gamma\frac{c}{\int c}\right]  .
\end{equation}

Now, taking into account eq. (\ref{teq_6}), we get that%
\begin{equation}
-\frac{c^{4}}{8\pi G\rho}\left[  \frac{2Ac}{\left(  \int c\right)  ^{3}}%
+\frac{8\pi\rho_{0}G}{c^{4}\left(  \int c\right)  ^{\gamma}}\left[
\frac{G^{\prime}}{G}-4\frac{c^{\prime}}{c}-\gamma\frac{c}{\int c}\right]
\right]  +\frac{G^{\prime}}{G}-4\frac{c^{\prime}}{c}=0,
\end{equation}
and hence we obtain%
\begin{equation}
G=\frac{A}{4\pi\rho_{0}\gamma}c^{4}\left(  \int c\right)  ^{\gamma-2},
\label{pili}%
\end{equation}
and in this way we find that
\begin{equation}
\Lambda=A\left(  1-\frac{2}{\gamma}\right)  \left(  \int c\right)  ^{-2}.
\label{enma}%
\end{equation}

As we can see, from eqs. (\ref{pili} and \ref{enma}), we have that are
verified the following relationships%
\begin{equation}
\frac{G\rho}{c^{4}}\thickapprox\left(  \int c\right)  ^{-2},\qquad
\Lambda\left(  \int c\right)  ^{2}=const.,
\end{equation}
in the next subsection (\ref{MC}), matter collineations approach, we will see
that these relationships are obtained in a trivial way.

Now, we will try to find the value of the constants $\left(  \alpha
_{i}\right)  _{i=1}^{3}.$ Taking into account the field eqs. (\ref{teq_1}%
-\ref{teq_4}) we find that, obviously eq. (\ref{teq_1}) vanish, but from eqs.
(\ref{teq_2}-\ref{teq_4}) we get%
\begin{align}
\alpha_{2}\left(  \alpha_{2}-1\right)  +\alpha_{3}\left(  \alpha_{3}-1\right)
+\alpha_{3}\alpha_{2}  &  =\frac{A}{\alpha}\left(  \alpha-2\right)
,\label{nsis1}\\
\alpha_{1}\left(  \alpha_{1}-1\right)  +\alpha_{3}\left(  \alpha_{3}-1\right)
+\alpha_{3}\alpha_{1}  &  =\frac{A}{\alpha}\left(  \alpha-2\right)
,\label{nsis2}\\
\alpha_{2}\left(  \alpha_{2}-1\right)  +\alpha_{1}\left(  \alpha_{1}-1\right)
+\alpha_{1}\alpha_{2}  &  =\frac{A}{\alpha}\left(  \alpha-2\right)  ,
\label{nsis3}%
\end{align}
where $A=\left(  \alpha_{1}\alpha_{2}+\alpha_{1}\alpha_{3}+\alpha_{2}%
\alpha_{3}\right)  ,$ $\alpha=\left(  \alpha_{1}+\alpha_{2}+\alpha_{3}\right)
.$ The system (\ref{nsis1}-\ref{nsis3}) has only two solutions%
\begin{align}
\alpha_{1}  &  =1-\alpha_{2}-\alpha_{3},\qquad\alpha_{2}=\alpha_{2}%
,\qquad\alpha_{3}=\alpha_{3},\qquad\text{and}\label{solnsys1}\\
\alpha_{1}  &  =\alpha_{2}=\alpha_{3}, \label{solnsys2}%
\end{align}
as we can see, the solution (\ref{solnsys1}) looks with physical meaning while
solution (\ref{solnsys2}) is the flat FRW one so in this case we must to rule
it out. This solution was obtained by Einstein\&de Sitter (\cite{EdS}) in 1932
for $\omega=0$, and later by Harrison (\cite{Harrison}) $\forall\omega.$ This
solution is quite similar to the obtained in the above section with the scale symmetry.

With regard to the solution (\ref{solnsys1}), it is also noted that the
solution $\alpha_{1}=1-\alpha_{2}-\alpha_{3},$ brings us to get $A=\left(
\alpha_{1}\alpha_{2}+\alpha_{1}\alpha_{3}+\alpha_{2}\alpha_{3}\right)
=\alpha_{2}+\alpha_{3}-\alpha_{2}^{2}-\alpha_{3}^{2}-\alpha_{2}\alpha_{3}>0,$
$\forall\alpha_{2},\alpha_{3}\in\left(  0,1\right)  .$ Note that the solution
(\ref{solnsys1}) verifies the relationship
\begin{equation}
\sum\alpha_{i}=1,\qquad\sum\alpha_{i}^{2}<1,
\end{equation}
this is the same class of solutions that we got in our previous paper
(\cite{tony1}) where we studied a Bianchi I model with $G$ and $\Lambda$
varying. \ See (\cite{tony1}) as well as the end of section \ref{KSS} for a
discussion of this class of solutions. Therefore we have found a similar
behavior as the obtained one in (\cite{HW}, we say similar because these
authors only study standard models i.e. models where the \textquotedblleft
constants\textquotedblright\ are true constants, in fact $\Lambda=0$), except
than here this result is valid for all equation of state i.e. $\forall\omega.$
Nevertheless in reference (\cite{griego}), the authors claim that the solution
must verify both conditions, i.e. $\sum\alpha_{i}=1=\sum\alpha_{i}^{2},$ (see
\cite{Kasner} \ and \ \cite{SH}).

We would like to stress that in this case, it is essential to take into
account the effects of a $c-$var into the field equations (as in this case).
For example, from eq. (\ref{teq_2})we find that:%
\begin{equation}
\alpha_{2}\frac{c^{\prime}}{\int c}+\alpha_{2}\left(  \alpha_{2}-1\right)
\left(  \frac{c}{\int c}\right)  ^{2}+\alpha_{3}\frac{c^{\prime}}{\int
c}+\alpha_{3}\left(  \alpha_{3}-1\right)  \left(  \frac{c}{\int c}\right)
^{2}-\left(  \alpha_{2}+\alpha_{3}\right)  \frac{c}{\int c}\frac{c^{\prime}%
}{c}+\alpha_{2}\alpha_{3}\left(  \frac{c}{\int c}\right)  ^{2}=\frac{A}%
{\alpha}\left(  \alpha-2\right)  \left(  \frac{c}{\int c}\right)  ^{2},
\end{equation}
simplifying, we get%
\begin{align}
\left[  \alpha_{2}\left(  \alpha_{2}-1\right)  +\alpha_{3}\left(  \alpha
_{3}-1\right)  +\alpha_{2}\alpha_{3}\right]  \left(  \frac{c}{\int c}\right)
^{2}  &  =\frac{A}{\alpha}\left(  \alpha-2\right)  \left(  \frac{c}{\int
c}\right)  ^{2},\\
\alpha_{2}\left(  \alpha_{2}-1\right)  +\alpha_{3}\left(  \alpha_{3}-1\right)
+\alpha_{2}\alpha_{3}  &  =\frac{A}{\alpha}\left(  \alpha-2\right)  ,
\end{align}
and so on.

But if we take the field equations in the usual way i.e.%
\begin{equation}
\frac{\ddot{Y}}{Y}+\frac{\ddot{Z}}{Z}+\frac{\dot{Z}}{Z}\frac{\dot{Y}}%
{Y}=-\frac{8\pi G}{c^{2}}\omega\rho+\Lambda c^{2},
\end{equation}
it yields%
\begin{align}
\alpha_{2}\left(  \frac{c^{\prime}}{\int c}+\left(  \alpha_{2}-1\right)
\left(  \frac{c}{\int c}\right)  ^{2}\right)  +\alpha_{3}\left(
\frac{c^{\prime}}{\int c}+\left(  \alpha_{3}-1\right)  \left(  \frac{c}{\int
c}\right)  ^{2}\right)  +\alpha_{2}\alpha_{3}\left(  \frac{c}{\int c}\right)
^{2}  &  =\frac{A}{\alpha}\left(  \alpha-2\right)  \left(  \frac{c}{\int
c}\right)  ^{2},\\
\alpha_{2}\left(  \alpha_{2}-1\right)  +\alpha_{3}\left(  \alpha_{3}-1\right)
+\alpha_{2}\alpha_{3}+\left(  \frac{c^{\prime}}{c}\frac{\int c}{c}\right)
\left(  \alpha_{2}+\alpha_{3}\right)   &  =\frac{A}{\alpha}\left(
\alpha-2\right)  .
\end{align}

An important observational quantity is the deceleration parameter $q=\frac
{d}{dt}\left(  \frac{1}{H}\right)  -1$. The sign of the deceleration parameter
indicates whether the model inflates or not. The positive sign of $q$
corresponds to \textquotedblleft standard\textquotedblright\ decelerating
models whereas the negative sign indicates inflation. Therefore we have%
\begin{equation}
H=\alpha\frac{c}{\int cdt}=\frac{c}{\int cdt},\qquad\text{and \qquad}%
q=\frac{d}{dt}\left(  \frac{1}{H}\right)  -1=-\frac{c^{\prime}}{c}\frac{\int
c}{c},
\end{equation}
note that $\alpha=\sum\alpha_{i}=1,$ furthermore we find that
\begin{equation}
\rho=\rho_{0}\left(  \int cdt\right)  ^{-(1+\omega)},\qquad G=\frac{A}%
{4\pi\rho_{0}\left(  1+\omega\right)  }c^{4}\left(  \int c\right)  ^{\omega
-1},\qquad\Lambda=A\left(  1-\frac{2}{(1+\omega)}\right)  \left(  \int
c\right)  ^{-2},
\end{equation}
so, depending on the choice of the function $c(t)$ we will have different
behaviors for each quantity but, we may impose some restrictions like the
following ones.

For the energy density, $\rho,$ since it must be a decreasing time function
for all $\omega\in(-1,1]$, we find that this is only possible iff $\left(
\int cdt\right)  $ is a growing time function. Note that if we consider the
case $c(t)=c_{0}=const.,$ then we have, $\left(  \int cdt=c_{0}t\right)  ,$
which is a growing time function. If $\omega<-1$ (phantom case), then, $\rho,$
is growing, but choosing a time decreasing $\int cdt,$ we may do that $\rho$
will be a time decreasing function as it is expected.

For $G,$ it is impossible to know beforehand which will be its behavior since
depends on $c,$ in the following way, $c^{4}\left(  \int c\right)  ^{\omega
-1},$ so we only may say that for $\omega=1,$ $G\thickapprox c^{4}$, while
$\forall\omega\in(-1,1),$ $\left(  \int c\right)  ^{\omega-1}$ is a decreasing
function on time, since we have impose that $\left(  \int cdt\right)  $ must
be a growing time function.

For $\Lambda,$ we may say that is a negative decreasing function on time
$\forall\omega\in(-1,1),$ but for $\omega=1,$ $\Lambda=0$ i.e. vanish, and for
$\omega<-1$ (phantom case) $\Lambda$ is positive as well as $\forall\omega>1$
(see for example \cite{Turok}). Note that $A>0.$ Furthermore, if $c=const.$
then it is regained all the results obtained in (\cite{tony1}), as for example
the relationships $G\rho\thickapprox t^{-2},$ and $\Lambda\thickapprox
t^{-2}.$ We would like to stress that this result, $\Lambda<0$ is not new in
the literature, see for example, T.Padmanabhan and S.M. Chitre, (\cite{padna}%
), nevertheless the recent observations suggest us that $\Lambda$ must be
positive (\cite{cc1}-\cite{cc4}), so in order to reconcile our results with
the observational data we need to consider that $\omega\in\left(
-\infty,1\right)  \cup\left(  1,\infty\right)  .$ With these values for the
equation of state it is observed that $G$ is growing if $\omega\in\left(
1,\infty\right)  $ (see \cite{Turok}) while $\rho$ is decreasing.

Since our model is formally self-similar, then (\cite{Jantzen}-\cite{Wainwrit}%
) have shown, that all the quantities must follow a power law, so, we may
assume that \ for example, $c$ takes the following form: $c(t)=c_{0}%
t^{\epsilon},$ with $\epsilon\in\mathbb{R}.$ Hence, we find from the
definition of the Hubble parameter and the deceleration parameter that
\begin{equation}
H=\frac{\epsilon+1}{t},\qquad q=-\frac{\epsilon}{\epsilon+1},
\end{equation}
and imposing the condition $q<0$ we find that $\epsilon\in\left(
0,\infty\right)  ,$ the special case, $\epsilon=-1,$ is forbidden, note that
$\int cdt=\frac{c_{0}}{\epsilon+1}t^{\epsilon+1}>0$ and a growing time
function$,\ \forall\epsilon\in\left(  0,\infty\right)  .$ So, from physical
considerations we find that%
\begin{equation}
\rho\thickapprox t^{-(1+\omega)\left(  \epsilon+1\right)  },\qquad
G\thickapprox t^{4\epsilon+(\omega-1)\left(  \epsilon+1\right)  }%
,\qquad\Lambda\thickapprox t^{-2\left(  \varepsilon+1\right)  },
\end{equation}
with $\epsilon\in\left(  0,\infty\right)  .$ We may also argue that since
$\Lambda=A\left(  1-\frac{2}{(1+\omega)}\right)  \left(  \int c\right)
^{-2},$ must be a decreasing time function, this is only possible iff
$\epsilon\in\left(  -1,\infty\right)  ,$ and therefore we find that
$\epsilon\in\left(  0,\infty\right)  .$ Therefore, if we take into account
these considerations, then we arrive to the conclusion that $c$ must be a
growing time functions, while $\Lambda$ is a decreasing time function and its
sing only depends on the equation of state. With regard to $G$, we may say
that its behavior depends on two parameters $\left(  \epsilon,\omega\right)
,$ so if $\epsilon\rightarrow0^{+}$ i.e. is a small positive number, and
$\omega\in\left(  -\infty,1\right)  $ then $G$ is a decreasing time function
but if $\omega\in\left(  1,\infty\right)  $ (see \cite{Turok}) then is growing
while if $\epsilon\rightarrow1$ and $\omega\in\left(  -1,\infty\right)  $ then
$G$ is a growing time function. Other possibilities could be considered
playing with different values of $\left(  \epsilon,\omega\right)  .$

Before ending, we would like to emphasize that, as it is observed, we have
choose, $\int cdt=\frac{c_{0}}{\varepsilon+1}t^{\varepsilon+1},$ instead of,
$\int cdt=\frac{c_{0}}{\varepsilon+1}t^{\varepsilon+1}+K,$ where $K$ is an
integrating constant. In this case $K=0,$ otherwise the resulting vector field
is not homothetic, i.e. it is not verified the eq. $L_{V}g=2g.$ If we fix,
$\epsilon=0,$ we regain the usual homothetic vector field i.e. eq.
(\ref{homoc}) (see (\cite{tony1})).

With regard to the curvature behavior we find that%
\begin{equation}
I_{1}=\frac{4}{\left(  \int c\right)  ^{4}}f(\alpha_{2},\alpha_{3}),\qquad
I_{2}=\frac{4}{\left(  \int c\right)  ^{4}}\left(  \alpha_{2}+\alpha
_{3}-\alpha_{2}^{2}-\alpha_{3}^{2}-\alpha_{2}\alpha_{3}\right)  ^{2}%
=\frac{4A^{2}}{\left(  \int c\right)  ^{4}},
\end{equation}
finding that if $\int c$ is a growing time function (as we have pointed out
above) then we get a singular behavior since $I_{1}$ and $I_{2}$ tend to
infinite as $t$ goes to zero.

The non-cero components of the Weyl tensor are:%
\begin{align}
C_{1212}  &  =-\frac{1}{3}c^{2}\tilde{A}\left(  \int c\right)  ^{-2\left(
\alpha_{2}+\alpha_{3}\right)  },\qquad C_{1313}=-\frac{1}{3}c^{2}B\left(  \int
c\right)  ^{-2\left(  1-\alpha_{2}\right)  },\qquad C_{1414}=-\frac{1}{3}%
c^{2}D\left(  \int c\right)  ^{-2\left(  1-\alpha_{3}\right)  },\nonumber\\
C_{2323}  &  =\frac{1}{3}c^{2}D\left(  \int c\right)  ^{-2\alpha_{3}},\qquad
C_{1313}=\frac{1}{3}c^{2}B\left(  \int c\right)  ^{-2\alpha_{2}},\qquad
C_{1414}=\frac{1}{3}c^{2}\tilde{A}\left(  \int c\right)  ^{-2\left(
1-\alpha_{2}-\alpha_{3}\right)  },
\end{align}
where $\tilde{A}=\left(  -\alpha_{2}-\alpha_{3}+\alpha_{2}^{2}+\alpha_{3}%
^{2}+4\alpha_{2}\alpha_{3}\right)  ,$ $B=\left(  -\alpha_{2}+2\alpha
_{3}+\alpha_{2}^{2}-\alpha_{3}^{2}-2\alpha_{2}\alpha_{3}\right)  ,$ and
$D=\left(  2\alpha_{2}-\alpha_{3}-2\alpha_{2}^{2}+\alpha_{3}^{2}-2\alpha
_{2}\alpha_{3}\right)  ,$ as above, we find that the Weyl tensor tends to
infinity if $\int c$ is a growing time function, note that $\alpha_{i}>0$
$\forall i.$

The non-cero components of the electric part are:%
\begin{equation}
E_{22}=-\frac{1}{3}\tilde{A}\left(  \int c\right)  ^{-2\left(  \alpha
_{2}+\alpha_{3}\right)  },\qquad E_{22}=-\frac{1}{3}B\left(  \int c\right)
^{-2\left(  1-\alpha_{2}\right)  },\qquad E_{22}=-\frac{1}{3}D\left(  \int
c\right)  ^{-2\left(  1-\alpha_{3}\right)  },
\end{equation}
finding that $E_{ij}\rightarrow\infty$ as $t\rightarrow0.$ Therefore the Weyl
invariant yields%
\begin{equation}
I_{3}=\frac{16}{3}\frac{f(\alpha_{2},\alpha_{3})}{\left(  \int c\right)  ^{4}%
},
\end{equation}
and the gravitational entropy is
\begin{equation}
P^{2}=\frac{I_{3}}{I_{2}}=\frac{4}{3}\left(  \frac{I_{1}}{I_{2}}-\frac{1}%
{3}\frac{R^{2}}{I_{2}}-2\right)  \neq0.
\end{equation}

So the obtained solution is singular.

\subsection{Matter collineations.\label{MC}}

In recent years, much interest has been shown in the study of matter
collineation (MCs) (see for example \cite{Sharif}-\cite{TA}). A vector field
along which the Lie derivative of the energy-momentum tensor vanishes is
called an MC, i.e.%
\begin{equation}
{\mathcal{L}}_{V}T_{ij}=0,
\end{equation}
where $V^{i}$ is the symmetry or collineation vector. The MC equations, in
component form, can be written as
\begin{equation}
T_{ij,k}V^{k}+T_{ik}V_{,j}^{k}+T_{kj}V_{,i}^{k}=0,
\end{equation}
where the indices $i,j,k$ run from $0$ to $3$. Also, assuming the Einstein
field equations, a vector $V^{i}$ generates an MC if ${\mathcal{L}}_{V}%
G_{ij}=0$. It is obvious that the symmetries of the metric tensor (isometries)
are also symmetries of the Einstein tensor $G_{ij}$, but this is not
necessarily the case for the symmetries of the Ricci tensor (Ricci
collineations) which are not, in general, symmetries of the Einstein tensor.
If $V$ is a Killing vector (KV) (or a homothetic vector), then ${\mathcal{L}%
}_{V}T_{ij}=0$, thus every isometry is also an MC but the converse is not
true, in general. Notice that collineations can be proper (non-trivial) or
improper (trivial). Proper MC is defined to be an MC which is not a KV, or a
homothetic vector.

Carot et al (see \cite{ccv}) and Hall et al.(see \cite{hrv}) have noticed some
important general results about the Lie algebra of MCs.

Let $M$ be a spacetime manifold. Then, generically, any vector field $V$ on
$M$ which simultaneously satisfies ${\mathcal{L}}_{V}T_{ab}=0$
($\Leftrightarrow{\mathcal{L}}_{V}G_{ab}=0$) and ${\mathcal{L}}_{V}C_{bcd}%
^{a}=0$ is a homothetic vector field.

If $V$ is a Killing vector (KV) (or a homothetic vector), then ${\mathcal{L}%
}_{V}T_{ab}=0$, thus every isometry is also an MC but the converse is not
true, in general. Notice that collineations can be proper (non-trivial) or
improper (trivial). Proper MC is defined to be an MC which is not a KV, or a
homothetic vector.

Since the ST is SS then we already know that the SS vector field is also
matter collineation i.e. we would like to explore how such symmetries allow us
to obtain relationships between the quantities in such a way that it is not
necessary to make any hypothesis to a solution to the field equations. In
order to do that we need to modify the usual MC equations since with the usual
one we are not able to obtain information about the behavior of $G,c$ and
$\Lambda.$ Therefore, following the same steeps as in ref (\cite{tony1}), will
be enough to check the following relationships:
\begin{equation}
L_{HO}\left(  \frac{G(t)}{c^{4}}T_{ij}\right)  =0.
\end{equation}
where $HO$ is given by eq. (\ref{HO1}).

In this case, we get from the resulting equations the following results:%
\begin{align}
\left(  \frac{G^{\prime}}{G}-4\frac{c^{\prime}}{c}+\frac{\rho^{\prime}}{\rho
}+\frac{2c}{\int cdt}\right)   &  =0,\qquad\Longleftrightarrow\qquad\frac
{G}{c^{4}}\rho\thickapprox\left(  \int cdt\right)  ^{-2},\label{m1}\\
\left(  -H_{1}+\frac{\int c}{c}\left(  H_{1}\frac{c^{\prime}}{c}-H_{1}%
^{\prime}\right)  \right)   &  =0,\qquad\Longleftrightarrow\qquad
\Longleftrightarrow X=X_{0}\left(  \int cdt\right)  ^{\alpha_{1}},\label{m2}\\
\text{similar result for }Y\text{ i.e. }Y  &  =Y_{0}\left(  \int cdt\right)
^{\alpha_{2}},\\
\text{similar result for }Z\text{ i.e. }Z  &  =Z_{0}\left(  \int cdt\right)
^{\alpha_{3}},\\
\left(  \frac{G^{\prime}}{G}-4\frac{c^{\prime}}{c}+\frac{p^{\prime}}{p}%
+\frac{2c}{\int cdt}\right)   &  =0,\qquad\Longleftrightarrow\qquad\frac
{G}{c^{4}}p\thickapprox\left(  \int cdt\right)  ^{-2}.
\end{align}

To end, in order to get \ information about the behavior of $\Lambda,$ we
consider the generalized MC eq., so we check again that:
\begin{equation}
L_{HO}\left(  \frac{G(t)}{c^{4}}T_{ij}-\Lambda(t)g_{ij}\right)  =0.
\end{equation}
finding the same result with regard to $(X,Y,Z),$ i.e. the scale factors as
well as for the energy density and the pressure, but the important
relationship here is the behavior of $\Lambda,$ where
\begin{align}
\left(  \frac{G^{\prime}}{G}-4\frac{c^{\prime}}{c}+\frac{\rho^{\prime}}{\rho
}+\frac{2c}{\int c}\right)   &  =-\frac{\Lambda c^{4}}{G\rho}\left(
\frac{\Lambda^{\prime}}{\Lambda}+2\frac{c}{\int c}\right)  ,\label{defm1}\\
\left(  -H_{1}+\frac{\int c}{c}\left(  H_{1}\frac{c^{\prime}}{c}-H_{1}%
^{\prime}\right)  \right)   &  =0,\qquad\Longleftrightarrow X=X_{0}\left(
\int cdt\right)  ^{\alpha_{1}}, \label{defm2}%
\end{align}
obtaining in this way%
\begin{equation}
\frac{G}{c^{4}}\rho\thickapprox\left(  \int cdt\right)  ^{-2},\qquad\text{and
\qquad}\Lambda\left(  \int cdt\right)  ^{2}=const.
\end{equation}
while if we fix $c=const$, (compare these results with the obtained ones in
(\cite{tony1})), then it is regained the usual relationship for the inertia as
well for the cosmological constant i.e.
\begin{equation}
\frac{G}{c^{2}}\rho\thickapprox t^{-2},\qquad\Lambda c^{2}=t^{-2}.
\end{equation}

As we have pointed out in the above section, all these result are verified by
the SS solution.

\section{Kinematic Self-similarity.\label{KSS}}

Kinematic self-similarity has been defined in the context of relativistic
fluid mechanics as an example of incomplete similarity (see for example
\cite{CH1}-\cite{Sintes-KSS}). It should be noted that the introduction of
incomplete similarity to general relativity is not unique.

A spacetime is said to be kinematic self-similar if it admits a kinematic
self-similar vector ${V}$ which satisfies the conditions
\begin{align}
{\mathcal{L}}_{V}h_{ij}  &  =2\delta h_{ij},\label{kss}\\
{\mathcal{L}}_{V}u_{i}  &  =\alpha u_{i}, \label{gss}%
\end{align}
where $u^{i}$ is the four-velocity of the fluid and $h_{ij}=g_{ij}+u_{i}u_{j}
$ is the projection tensor, and $\alpha$ and $\delta$ are constants~.

If $\delta\neq0$, the similarity transformation is characterized by the
scale-independent ratio $\alpha/\delta$, which is referred to as the
similarity index. If the ratio is unity, ${V}$ turns out to be a homothetic
vector. In the context of kinematic self-similarity, homothety is referred to
as self-similarity of the first kind. If $\alpha=0$ and $\delta\neq0$, it is
referred to as self-similarity of the zeroth kind. If the ratio is not equal
to zero or one, it is referred to as self-similarity of the second kind. If
$\alpha\neq0$ and $\delta=0$, it is referred to as self-similarity of the
infinite kind. If $\delta=\alpha=0$, ${V}$ turns out to be a Killing vector.

From the Einstein equation (\ref{einstein}), we can derive
\begin{equation}
\mathcal{L}_{V}G_{ij}=8\pi G\mathcal{L}_{V}T_{ij}, \label{intcondition}%
\end{equation}
this equation is called the integrability condition.

When a perfect fluid is irrotational, i.e., $\omega_{ij}=0$, the Einstein
equations and the integrability conditions (\ref{intcondition}) give%
\begin{equation}
(\alpha-\delta)\mathcal{R}_{ij}=0,
\end{equation}
where $\mathcal{R}_{ij}$ is the Ricci tensor on the hypersurface orthogonal to
$u^{i}$. This means that if a solution is kinematic self-similar but not
homothetic and if the fluid is irrotational, then the hypersurface orthogonal
to fluid flow is flat.

From the physical point of view the detailed study of cosmological models
admitting KSS shows that they can represent asymptotic states of more general
models or, under certain conditions, they are asymptotic to an exact
homothetic solution \cite{Coley-KSS,Benoit-Coley}.

Therefore and following the same idea as in the above sections we would like
to extend this hypothesis in order to find exact solutions to cosmological
models with time varying constant.

Kinematic self-similarity is characterized by the equations (\ref{kss}%
-\ref{gss}), so in this way it is found that the vector field $V:=KSS$ is:%
\begin{equation}
V=\left(  -\alpha\frac{\int cdt}{c}\right)  \partial_{t}+f_{1}x\partial
_{x}+f_{2}y\partial_{y}+f_{3}z\partial_{z},
\end{equation}
where%
\begin{equation}
f_{1}=\left(  \delta+\left(  \alpha\frac{\int cdt}{c}\right)  \frac{\dot{X}%
}{X}\right)  ,\quad f_{2}=\left(  \delta+\left(  \alpha\frac{\int cdt}%
{c}\right)  \frac{\dot{Y}}{Y}\right)  ,\quad f_{3}=\left(  \delta+\left(
\alpha\frac{\int cdt}{c}\right)  \frac{\dot{Z}}{Z}\right)  .
\end{equation}

As in the case of the homothetic vector field in this case it is necessary to
satisfy the following ODE
\begin{equation}
\left(  \alpha\frac{\int cdt}{c}\right)  H_{1}^{\prime}=-\alpha H_{1}\left(
1-\frac{c^{\prime}}{c}\frac{\int cdt}{c}\right)  ,\qquad\Longrightarrow\qquad
H_{1}=a_{1}\left(  \frac{\int cdt}{c}\right)  ^{-1},
\end{equation}
arriving to the same conclusion as in the SS solution i.e.%
\begin{equation}
H_{1}=\frac{X_{1}^{\prime}}{X_{1}}\qquad\Longrightarrow\qquad X_{1}%
=X_{0}\left(  \int cdt\right)  ^{a_{1}}.
\end{equation}

In this way and following the same procedure as in the above section, we find
that%
\begin{equation}
H=a\left(  \frac{c}{\int c}\right)  ,\text{ \ with}\qquad a=\sum_{i=1}%
^{3}a_{i}, \label{olga}%
\end{equation}
and therefore
\begin{equation}
\rho=\rho_{0}\left(  \int cdt\right)  ^{-a\left(  \omega+1\right)  },
\end{equation}
as it is observed if we choose the case $c=const.$ then we regain the usual
results as it is expected.

The shear behaves as:%
\begin{equation}
\sigma^{2}=\frac{1}{3}\left(  \sum_{i=1}^{3}\alpha_{i}^{2}-\sum_{i\neq
j}\alpha_{i}\alpha_{j}\right)  \left(  \frac{1}{\int cdt}\right)  ^{2}.
\end{equation}

From the field eq. (\ref{teq_1}) we get%
\begin{equation}
\psi=\frac{8\pi G}{c^{4}}\rho+\Lambda,\qquad\text{with}\qquad\psi=A\left(
\frac{1}{\int c}\right)  ^{2}, \label{olga1}%
\end{equation}
where $A=\left(  \alpha_{1}\alpha_{2}+\alpha_{1}\alpha_{3}+\alpha_{2}%
\alpha_{3}\right)  ,$ and $\gamma=a\left(  \omega+1\right)  ,$ and $a$ is
given by eq. (\ref{olga}), so, from eq. (\ref{olga1}) we get%
\begin{equation}
\Lambda^{\prime}=\psi^{\prime}-8\pi\frac{G}{c^{4}}\rho\left[  \frac{G}%
{G}+\frac{\rho^{\prime}}{\rho}-4\frac{c^{\prime}}{c}\right]  .
\end{equation}

Now, taking into account eq. (\ref{teq_6}), we get that%
\begin{equation}
\Lambda^{\prime}=\psi^{\prime}-8\pi\frac{G}{c^{4}}\rho\left[  \frac
{\rho^{\prime}}{\rho}-\frac{\Lambda^{\prime}c^{4}}{8\pi G\rho}\right]
,\qquad\Longrightarrow\qquad G=\frac{\psi^{\prime}c^{4}}{8\pi\rho^{\prime}},
\end{equation}
simplifying it, we obtain%
\begin{equation}
G=\frac{2A}{8\pi\gamma\rho_{0}}\frac{c^{4}}{\left(  \int c\right)  ^{2-\gamma
}},
\end{equation}
note that if $c=const.$then we get%
\begin{equation}
G=\frac{2A}{8\pi\gamma\rho_{0}}\frac{c^{2}}{\left(  t\right)  ^{2-\gamma}}.
\end{equation}

In this way we find from eq. (\ref{olga1}) that
\begin{equation}
\Lambda=\frac{A\left(  \gamma-2\right)  }{\gamma}\left(  \int c\right)  ^{-2},
\end{equation}
regaining the usual expression when $c=const.$%
\begin{equation}
\Lambda c^{2}=\frac{A\left(  \gamma-2\right)  }{\gamma}\frac{1}{t^{2}}.
\end{equation}

In this way we find that%
\begin{align}
\alpha_{2}\left(  \alpha_{2}-1\right)  +\alpha_{3}\left(  \alpha_{3}-1\right)
+\alpha_{3}\alpha_{2}  &  =\frac{A}{\alpha}\left(  \alpha-2\right)
,\label{nsis11}\\
\alpha_{1}\left(  \alpha_{1}-1\right)  +\alpha_{3}\left(  \alpha_{3}-1\right)
+\alpha_{3}\alpha_{1}  &  =\frac{A}{\alpha}\left(  \alpha-2\right)  ,\\
\alpha_{2}\left(  \alpha_{2}-1\right)  +\alpha_{1}\left(  \alpha_{1}-1\right)
+\alpha_{1}\alpha_{2}  &  =\frac{A}{\alpha}\left(  \alpha-2\right)  ,
\label{nsis31}%
\end{align}
finding that this is the same system of equations that we had in the SS
solution and therefore we get the same set of solutions i.e. the system
(\ref{nsis11}-\ref{nsis31}) has only two solutions%
\begin{align}
\alpha_{1}  &  =1-\alpha_{2}-\alpha_{3},\qquad\alpha_{2}=\alpha_{2}%
,\qquad\alpha_{3}=\alpha_{3},\qquad\text{and}\label{solnsis11}\\
\alpha_{1}  &  =\alpha_{2}=\alpha_{3}, \label{solnsis22}%
\end{align}

Hence we arrive to the same conclusions as in the SS solution, i.e., solution
(\ref{solnsis11}) looks with physical meaning while solution (\ref{solnsis22})
is the flat FRW one, so in this case we must to rule it out. Therefore we
arrive to the following result%
\begin{equation}
\sum\alpha_{i}=1,\qquad\sum\alpha_{i}^{2}<1.
\end{equation}

Before ending, we would like to emphasize that, in this case, we may choose,
$\int cdt=\frac{c_{0}}{\varepsilon+1}t^{\varepsilon+1}+K,$ where $K$ is an
integrating constant, $K\neq0,$ in such a way that the resulting solution is
non-singular, and it is quite similar to the obtained one in the case of the
full symmetry obtained in section (\ref{liefull}). If we fix, $\epsilon=0,$ we
regain the usual kinematical self-similar vector field
\begin{equation}
KSS=-(\alpha t+\beta)\partial_{t}+f_{1}x\partial_{x}+f_{2}y\partial_{y}%
+f_{3}z\partial_{z},
\end{equation}
where%
\begin{equation}
f_{1}=\left(  \delta+(\alpha t+\beta)\frac{\dot{X}}{X}\right)  ,\quad
f_{2}=\left(  \delta+(\alpha t+\beta)\frac{\dot{Y}}{Y}\right)  ,\quad
f_{3}=\left(  \delta+(\alpha t+\beta)\frac{\dot{Z}}{Z}\right)  ,
\end{equation}
see (\cite{tony1}) for details.

Note that the solution (\ref{solnsys1}) does not verify the relationship,
$\sum\alpha_{i}^{2}=1,$i.e. it is Kasner's type (see \cite{Kasner}, \cite{SH}
and in particular \cite{griego}). But, if for example we suppose that solution
(\ref{solnsys1}) verifies the conditions%
\begin{equation}
\sum\alpha_{i}=1=\sum\alpha_{i}^{2}%
\end{equation}
this means that%
\begin{equation}
\alpha_{1}+\alpha_{2}+\alpha_{3}=1,\qquad\left(  -\alpha_{2}-\alpha_{3}%
+\alpha_{2}^{2}+\alpha_{3}^{2}+\alpha_{2}\alpha_{3}\right)  =0, \label{enma11}%
\end{equation}
and therefore%
\begin{equation}
\alpha_{1}=\frac{1}{2}\left(  1-\alpha_{3}-\sqrt{1+2\alpha_{3}-3\alpha_{3}%
^{2}}\right)  ,\quad\alpha_{2}=\frac{1}{2}\left(  1-\alpha_{3}+\sqrt
{2\alpha_{3}-3\alpha_{3}^{2}+1}\right)  ,\quad\alpha_{3}=\alpha_{3},
\end{equation}
which is not a physical solution since not all the $\left(  \alpha_{i}\right)
\in\left(  0,1\right)  ,$ for example $\alpha_{1}\in\left(  -1,0\right)  .$
Furthermore $A=\left(  \alpha_{1}\alpha_{2}+\alpha_{1}\alpha_{3}+\alpha
_{2}\alpha_{3}\right)  =\alpha_{2}+\alpha_{3}-\alpha_{2}^{2}-\alpha_{3}%
^{2}-\alpha_{2}\alpha_{3},$ this means from eq. (\ref{enma11}) that $A$ is
equal nought i.e. $\left(  A=0\right)  $. Therefore%
\begin{equation}
\rho=\rho_{0}\left(  \int cdt\right)  ^{-\left(  \omega+1\right)  },\qquad
G=\frac{2A}{8\pi\gamma\rho_{0}}\frac{c^{4}}{\left(  \int c\right)  ^{2-\gamma
}}=0,\qquad\Lambda=\frac{A\left(  \gamma-2\right)  }{\gamma}\left(  \int
c\right)  ^{-2}=0,
\end{equation}
as it is expected for a vacuum solution. \ See for example A. Harvey
\cite{Harvey} for a review of Bianchi I solutions (Kasner-like solutions)

Furthermore, as we shown in (\cite{tony1}) this class of solution has
pathological curvature behavior since if $I_{2}=0,$ then the gravitational
entropy is infinite i.e. $P^{2}=\infty.$

\section{Conclusions.\label{Conclu}}

We have shown how to attack a perfect fluid Bianchi I with time varying
constants under the condition $\operatorname{div}T=0,$ and taking into account
the effects of a $c-var$ into the curvature tensor i.e. modifying the usual FE.

With the first of the exposed tactics, i.e. the Lie group one, we have solved
the field equations, solving only one ODE, eq. (\ref{defeq}), studying the
possible forms that take $G(t),$ $c(t)$ and $\Lambda(t),$ in order to make
\ eq. (\ref{dorota1}) integrable. We have started imposing a particular
symmetry, $X=(at+e)\partial_{t}+b\partial\rho,$ which as we already know
brings us to get power law solutions. To study all the possible symmetries
would result a very tedious work.

In this way we have obtained three exact solutions in function of the behavior
of $G(t),$ $c(t)$ and $\Lambda(t).$ In this case we have not been able to rule
out neither of them as in our previous work \cite{tony1}, where some of them
had, $\sigma=0$, i.e. the shear vanish, and therefore we have rejected this
solution since we are only interested in solutions that verify the condition
$\sigma\neq0.$ Here the situation is a bit complicated since all the solutions
depend of many integrating constants so it is really difficult to rule out
some of them as well as to determine their behavior. \ Nevertheless, when we
calculate the numerical values of the exponents of the scale factors $\left(
\alpha_{i}\right)  _{i=1}^{3},$ we have shown that the only possible solution
is the flat FRW one, but, at this time, with $G,c$ and $\Lambda$ time varying.
This has been a really surprising result, since we think that the followed
tactic, i.e. solving eq. (\ref{dorota1}) without imposing any assumption ad
hoc, brings us to get consistent results in the framework of Bianchi I models
i.e. a solution with $\sigma\neq0.$ In this way we have arrived to the same
solutions as the obtained ones in \cite{TonyCarames}, as well as to the same
scenario as in our previous paper \cite{tony1}. We think that the followed
tactic is too restrictive, for this reason we are only able to obtain this
class of solutions. Nevertheless there are other Lie approaches as the
followed by M. Szydlowski et al (see \cite{Marek}) which we think that may be
more useful than the followed one here. As we will show in appendix
\ref{appA}, if we try to improve the obtained solutions through the study of a
third order ODE through the LM, we arrive to the same solutions and hence to
the same conclusions. Therefore, since there are many constrains, then we are
introducing several integrating constants which add uncertain to the obtained
solutions and hence we are not able to improve the obtained solution
integrating the second order ODE, we only obtain the same order of magnitude
in each quantity, that's all. As we have mentioned above, in appendix
\ref{appB} we will study a third order ODE which has been obtained without the
assumption of $c-$var affecting to the curvature tensor. We arrive to the same
solutions as the obtained ones in appendix \ref{appA}, and therefore we
conclude that at least in order of magnitude, there is no difference between
both approaches.

At the same time we have shown that it is not necessary to make any ad hoc
assumption or to take into account any previous hypothesis or considering any
hypothetical behavior for any quantity since all these hypotheses could be
deduced from the symmetry principles, as for example using the Lie group
methods or studying the model from the point of view of the geometrical
symmetries i.e. SS etc...

With regard to the others employed tactics to study the field equations, i.e.
SS, MC and KSS, we have shown that both tactics are quite similar and that
they bring us to get really similar results, actually as we already know, with
the SS and the MC we get the same results.

We have shown that the solution obtained with the SS hypothesis is also quite
similar to the obtained one using the Lie method under the scale symmetry. In
fact we have got two solutions, the flat FRW one and a Kasner-like solution.

Since in this case, all the obtained solutions, for each quantity, depend on
$\left(  \int c(t)dt\right)  ,$ it is difficult to determine the behavior of
each quantity. Nevertheless, we are able to arrive to some conclusions under
the hypothesis $q<0,$ (where $q$ stands for the deceleration parameter) which
are that $c$ must be a growing time functions while $\Lambda$ is a decreasing
time function and whose sign depends on the equation of state, finding that we
only get a positive cosmological constant if $\omega\in\left(  -\infty
,-1\right)  \cup\left(  1,\infty\right)  $. With regard to $G$, we may say
that its behavior depends on two parameters $\left(  \epsilon,\omega\right)
,$ so $G$ may be a decreasing time function as well as a growing time function
depending on the values of $\left(  \epsilon,\omega\right)  $. In the same way
as in \cite{tony1} we conclude that the exponents of the scale factor must
satisfy the conditions $\sum_{i=1}^{3}\alpha_{i}=1$ and $\sum_{i=1}^{3}%
\alpha_{i}^{2}<1,$ $\forall\omega,$ i.e. valid for all value of the equation
of state$,$ relaxing in this way the Kasner conditions.

We furthermore have pointed out, as it is well known, that if the ST is SS
then there is a vector field, $V\in\mathfrak{X}(M)$ that satisfies the
equation $\mathcal{L}_{V}g=2g,$ then such vector field must satisfy the
equation $\mathcal{L}_{V}T=0,$ i.e. a homothetic vector field is also a MC
vector field. In this occasion we only check that the homothetic vector field
verifies the reformulated MC equations (see \cite{tony1} for details) in order
to get information on the behavior of $G,c$ and $\Lambda$, arriving to the
same conclusions as in the SS section. Therefore we have shown that this
tactic would be very useful in the study of more complicated models as for
example the viscous ones.

With regard to the KSS solution, we have shown that it behaves like the SS
one, except that in this case, we obtain a non-singular behavior. We also have
show, that if one gets Kasner-like solutions i.e. they are verified the
conditions $\sum\alpha_{i}=1,$ and $\sum\alpha_{i}^{2}=1,$ then this class of
solutions bring us to get vanishing quantities i.e. $G=\Lambda=0,$ as well as
of obtaining a pathological curvature behavior since the model is Ricci flat
which means that $I_{2}=0,$ so the gravitational entropy is infinite.

\begin{acknowledgments}
I would like to thank T. Harko for his valuable comments and M. Szydlowsky for
drawing my attention on his work about Lie groups. Finally, to the anonymous
referee for her/his comments, which have contributed to improve the final
version of this work.
\end{acknowledgments}

\appendix

\section{Study of eq. (27).\label{appA}}

In section \ref{Lie}, we have studied a second order ODE and as we have been
able to see, since all the quantities depend of many integrating constants,
all the obtained solutions are very imprecise, i.e. they do not allow us to
know which is the real behavior of each quantity and therefore it is
impossible to rule them out, as in our previous paper \cite{tony1}. Since in
\cite{tony1} was very useful to study the third order ODE in this appendix we
would like to improve the obtained solutions in section \ref{Lie}. But,
unfortunately, as we will show, in this case we are not able to improve such
solutions obtaining only the same order of magnitude for each quantity.
Nevertheless, this study will allow us to show that there is no difference
between to study the resulting FE with $c-$var affecting to the curvature
tensors and the usual FE with $c-$var non-affecting to the curvature tensors.

Therefore, in this section we will study eq. (\ref{defeq}) through the Lie
group method. In particular we seek the forms of $G(t),c(t)$ and $\Lambda(t) $
for which our field equations admit symmetries i.e. they are integrable. Note
that this ODE has been obtained under the assumption that $c $ affects to the
curvature tensor i.e. $c-$var introduce some modifications into the curvature
tensor. So the equation under study is%
\begin{equation}
\dddot{\rho}=K_{1}\ddot{\rho}\frac{\dot{\rho}}{\rho}-K_{2}\frac{\dot{\rho}%
^{3}}{\rho^{2}}+\frac{G\rho^{2}}{c^{2}}\left[  K_{3}\frac{G^{\prime}}{G}%
-K_{4}\frac{\rho^{\prime}}{\rho}-K_{5}\frac{c^{\prime}}{c}\right]  -K_{6}\rho
cc^{\prime}\Lambda+\dot{\rho}\left(  \frac{c^{\prime\prime}}{c}-\frac
{c^{\prime2}}{c^{2}}\right)  +\frac{c^{\prime}}{c}\left(  \ddot{\rho}%
-\frac{\dot{\rho}^{2}}{\rho}\right)  ,
\end{equation}
where, $\left(  K_{i}\right)  _{i=1}^{6},$ are given by eqs. (\ref{choren0}).

Following the standard procedure we need to solve the next system of PDEs:%
\begin{align}
c^{4}\rho^{3}\xi_{\rho}  &  =0,\label{lie1}\\
c^{4}\rho^{3}\xi_{\rho\rho}  &  =0,\label{lie2}\\
K_{1}c^{4}\rho\eta-K_{1}c^{4}\rho^{2}\eta_{\rho}-3c^{3}\rho^{3}\xi_{\rho
}c^{\prime}-9c^{4}\rho^{3}\xi_{t\rho}+3c^{4}\rho^{3}\eta_{\rho\rho}  &
=0,\label{lie3}\\
-K_{1}c^{4}\rho^{2}\eta_{t}+\rho^{3}c^{4}\left[  \xi\left(  -\frac
{c^{\prime\prime}}{c}+\frac{c^{\prime2}}{c^{2}}\right)  -\xi_{t}%
\frac{c^{\prime}}{c}\right]  +3c^{4}\rho^{3}\eta_{t\rho}-3c^{4}\rho^{3}%
\xi_{tt}  &  =0,\label{lie4}\\
K_{1}c^{4}\rho^{2}\xi_{\rho\rho}+K_{2}c^{4}\rho\xi_{\rho}-c^{4}\rho^{3}%
\xi_{\rho\rho\rho}  &  =0,\label{lie5}\\
2K_{2}c^{4}\rho\eta_{\rho}-K_{1}c^{4}\rho^{2}\eta_{\rho\rho}+2c^{3}\rho^{2}%
\xi_{\rho}c^{\prime}+2K_{1}c^{4}\rho^{2}\xi_{t\rho}-2K_{2}c^{4}\eta-3c^{4}%
\rho^{3}\xi_{t\rho\rho}+c^{4}\rho^{3}\eta_{\rho\rho\rho}+c^{3}\rho^{3}%
\xi_{\rho\rho}c^{\prime}  &  =0, \label{lie6}%
\end{align}%
\[
-3c^{4}\rho^{3}\xi_{tt\rho}+3c^{4}\rho^{3}\eta_{t\rho\rho}+\rho^{2}%
c^{4}\left[  \xi\left(  \frac{c^{\prime\prime}}{c}-\frac{c^{\prime2}}{c^{2}%
}\right)  +\frac{c^{\prime}}{c}\left(  \eta_{\rho}+\xi_{t}-\frac{\eta}{\rho
}\right)  \right]  +3K_{2}c^{4}\rho\eta_{t}+
\]%
\begin{equation}
+3K_{4}c^{2}\rho^{4}G\xi_{\rho}-3c^{3}\rho^{3}\xi_{\rho}\left(  c^{\prime
\prime}+c^{\prime2}\right)  -2K_{1}c^{4}\rho^{2}\eta_{t\rho}+K_{1}c^{4}%
\rho^{2}\xi_{tt}+c^{3}\rho^{3}c^{\prime}\left(  2\xi_{t\rho}-\eta_{\rho\rho
}\right)  =0, \label{lie7}%
\end{equation}%
\begin{align*}
&  3c^{4}\rho^{3}\eta_{tt\rho}-c^{4}\rho^{3}\xi_{tt\,t}+K_{4}c^{2}\rho
^{4}G\left[  \xi\left(  \frac{G^{\prime}}{G}-2\frac{c^{\prime}}{c}\right)
+\frac{\eta}{\rho}+2\xi_{t}\right]  +\\
&  \rho^{3}c^{4}\left[  \xi\left(  3\frac{c^{\prime\prime}}{c}\frac{c^{\prime
}}{c}-\frac{c^{\prime\prime\prime}}{c}-2\frac{c^{\prime3}}{c^{3}}\right)
+2\xi_{t}\left(  \frac{c^{\prime2}}{c^{2}}-\frac{c^{\prime\prime}}{c}\right)
\right]  +2c^{3}\rho^{2}\eta_{t}c^{\prime}-K_{1}c^{4}\rho^{2}\eta_{tt}-
\end{align*}%
\begin{equation}
-2c^{3}\rho^{3}c^{\prime}\eta_{t\rho}+c^{3}\rho^{3}\xi_{tt}c^{\prime}%
-4K_{3}c^{2}\rho^{5}G^{\prime}\xi_{\rho}+4K_{5}cc^{\prime}\rho^{5}G\xi_{\rho
}+4K_{6}c^{5}c^{\prime}\rho^{4}\Lambda\xi_{\rho}=0, \label{lie8}%
\end{equation}%
\[
c^{4}\rho^{3}\eta_{tt\,t}-c^{3}\rho^{3}c^{\prime\prime}\eta_{t}+c^{2}\rho
^{3}c^{\prime2}\eta_{t}-c^{3}\rho^{3}c^{\prime}\eta_{tt}+K_{4}c^{2}\rho
^{4}G\eta_{t}+
\]%
\[
c^{2}\rho^{5}GK_{3}\left(  \xi\left(  2\frac{c^{\prime}}{c}\frac{G^{\prime}%
}{G}-\frac{G^{\prime\prime}}{G}\right)  +\frac{G^{\prime}}{G}\left(
\eta_{\rho}-2\frac{\eta}{\rho}-3_{t}\right)  \right)  +
\]%
\[
c^{2}\rho^{5}GK_{5}\left(  \xi\left(  \frac{c^{\prime}}{c}\frac{G^{\prime}}%
{G}-3\frac{(c^{\prime})^{2}}{c^{2}}+\frac{c^{\prime\prime}}{c}\right)
+\frac{c^{\prime}}{c}\left(  2\frac{\eta}{\rho}-\eta_{\rho}+3\xi_{t}\right)
\right)  +
\]%
\begin{equation}
K_{6}c^{6}\rho^{4}\Lambda\left[  \xi\left(  \frac{(c^{\prime})^{2}}{c^{2}%
}+\frac{c^{\prime\prime}}{c}+\frac{c^{\prime}}{c}\frac{\Lambda^{\prime}%
}{\Lambda}\right)  +\frac{c^{\prime}}{c}\left(  \frac{\eta}{\rho}-\eta_{\rho
}+3\xi_{t}\right)  \right]  =0. \label{lie9}%
\end{equation}

Imposing the symmetry $X=\left(  at+e\right)  \partial_{t}+b\rho\partial
_{\rho},$ i.e. $\left(  \xi=at+e,\,\eta=b\rho\right)  ,$ where $a,b,e\in
\mathbb{R}.$ Note that $\left[  a\right]  =\left[  b\right]  =1,$ i.e. they
are dimensionless constants but $\left[  e\right]  =T$, with respect to a
dimensional base $B=\left\{  L,M,T\right\}  ,$ we get the following
restrictions for $G(t),c(t)$ and $\Lambda(t)$.

From eq. (\ref{lie4}) we get%
\begin{equation}
\frac{c^{\prime\prime}}{c^{\prime}}-\frac{c^{\prime}}{c}=-\frac{a}{at+e}
\label{rest0}%
\end{equation}

From eq. (\ref{lie8}) we find that
\begin{equation}
\frac{G^{\prime}}{G}-2\frac{c^{\prime}}{c}=-\frac{b+2a}{at+e}, \label{rest1}%
\end{equation}
and
\begin{equation}
\left(  3\frac{c^{\prime\prime}}{c}-\frac{c^{\prime\prime\prime}}{c^{\prime}%
}-2\frac{c^{\prime2}}{c^{2}}\right)  =-2\left(  \frac{a}{at+e}\right)  ^{2},
\label{helen1}%
\end{equation}
where the most general solution for (\ref{helen1}) is
\begin{equation}
c=K_{0}\exp\left(  -\frac{K_{1}}{9}t^{3}+\frac{K_{2}}{9}\right)  t^{K_{2}/3},
\label{solhelen1}%
\end{equation}
where $\left(  K_{i}\right)  _{i=0}^{2}\in\mathbb{R},$ so a solution of
(\ref{rest0}) is a particular solution of (\ref{helen1}).

From eq. (\ref{lie9}) it is obtained:%
\begin{align}
\left(  at+e\right)  \left(  2\frac{c^{\prime}}{c}\frac{G^{\prime}}{G}%
-\frac{G^{\prime\prime}}{G}\right)  -\left(  b+3a\right)  \frac{G^{\prime}%
}{G}  &  =0,\label{rest2}\\
\left(  at+e\right)  \left(  \frac{c^{\prime}}{c}\frac{G^{\prime}}{G}%
-3\frac{(c^{\prime})^{2}}{c^{2}}+\frac{c^{\prime\prime}}{c}\right)
+\frac{c^{\prime}}{c}\left(  b+3a\right)   &  =0,\label{rest3}\\
\left(  at+e\right)  \left(  \frac{(c^{\prime})^{2}}{c^{2}}+\frac
{c^{\prime\prime}}{c}+\frac{c^{\prime}}{c}\frac{\Lambda^{\prime}}{\Lambda
}\right)  +3a\frac{c^{\prime}}{c}  &  =0, \label{rest4}%
\end{align}

Now from (\ref{rest3}) we get:%
\begin{equation}
\frac{G^{\prime}}{G}-3\frac{c^{\prime}}{c}+\frac{c^{\prime\prime}}{c^{\prime}%
}=-\frac{3a+b}{at+e}, \label{rest3-1}%
\end{equation}
and taking into account eq. (\ref{rest1}) we find that%
\begin{equation}
\frac{c^{\prime\prime}}{c^{\prime}}-\frac{c^{\prime}}{c}=-\frac{a}{at+e}.
\label{rest3-2}%
\end{equation}

In the same way, form eq. (\ref{rest4}) it is found that
\begin{equation}
\frac{c^{\prime}}{c}+\frac{c^{\prime\prime}}{c^{\prime}}+\frac{\Lambda
^{\prime}}{\Lambda}=-\frac{3a}{at+e}, \label{rest4-1}%
\end{equation}
and taking into account eq., (\ref{rest3-2}) we get:%
\begin{equation}
\frac{\Lambda^{\prime}}{\Lambda}+2\frac{c^{\prime}}{c}=-\frac{2a}{at+e}.
\label{rest4-2}%
\end{equation}

Therefore the restrictions (\ref{rest1}, \ref{rest3-2} and \ref{rest4-2} )
will be enough to find a solution for eq. (\ref{defeq}). As we can see, these
restrictions are the same than the obtained ones in section (\ref{Lie}, see
eqs. (\ref{silv1}-\ref{silv3})) where we studied the second order ODE and
therefore we expect to obtain a very similar result, we will find only a
little differences in the numerical constants but not in the order of
magnitude of each quantity. We may check how works these restrictions in the
case of the scale symmetry, since the rest of solutions will be obtained
copying the same steeps as the followed ones in section (\ref{Lie}).

\subsection{Scale symmetry.}

Making $e=0$ $i.e.$ considering only $\left(  \xi=at,\ \ \eta=b\rho\right)  ,$
we have to integrate eqs. (\ref{rest3-2}, \ref{rest1} and \ref{rest4-2}), so
\begin{align}
\frac{c^{\prime\prime}}{c^{\prime}}-\frac{c^{\prime}}{c}  &  =-\frac{1}%
{t},\label{elena}\\
\frac{G^{\prime}}{G}-2\frac{c^{\prime}}{c}  &  =-\frac{b+2a}{at}%
,\qquad\Longrightarrow\qquad\frac{G}{c^{2}}=Bt^{-\left(  2+\frac{b}{a}\right)
}\\
\frac{\Lambda^{\prime}}{\Lambda}+2\frac{c^{\prime}}{c}  &  =-\frac{2}%
{t},\qquad\Longrightarrow\qquad\Lambda c^{2}=\tilde{B}t^{-2},
\end{align}
$B,\tilde{B}\in\mathbb{R},$ therefore we get
\begin{align}
c  &  =c_{0}t^{c_{1}},\qquad c_{1},c_{0}\in\mathbb{R},\label{cscale}\\
G  &  =G_{0}t^{2(c_{1}-1-\frac{b}{2a})},\qquad G_{0}\in\mathbb{R}%
^{+},\label{Gscale}\\
\Lambda &  =\Lambda_{0}t^{-2(c_{1}+1)},\qquad\Lambda_{0}\in\mathbb{R},
\label{Lscale}%
\end{align}
where we assume that $G_{0}>0.$ Note that the obtained solution for $c(t)$
obviously verifies eq. (\ref{elena}) as well as it does verify eq.
(\ref{helen1}) but the most general solution of eq. (\ref{helen1}) i.e. eq.
(\ref{solhelen1}) does not verify eq. (\ref{elena}).

The invariant solution for the energy density is: $\rho=\rho_{0}t^{b/a},$ and
for physical reasons we impose the condition, $ab<0$ then $b<0.$ If we make
that this solution verifies eq. (\ref{defeq}) with $c(t),G(t)$ and
$\Lambda(t)$ given by eqs. (\ref{cscale}-\ref{Lscale}), we find the value of
constant $\rho_{0},$ so%
\begin{equation}
\rho_{0}=-\left[  \frac{c_{0}^{2}\left(  b^{2}+ab\left(  1+\omega\right)
(c_{1}+1)+3c_{0}^{2}c_{1}\Lambda_{0}a^{2}\left(  1+\omega\right)  ^{2}\right)
}{12\pi aG_{0}\left(  1+\omega\right)  ^{2}\left(  a\left(  2c_{1}%
+1+\omega\right)  +b\right)  }\right]  , \label{ayla1}%
\end{equation}
with the only restriction $\omega\neq-1,$ compare with eq. (\ref{ayla})$.$
Note that $ab<0,$ so we need to choice constants $\left(  c_{1},c_{0}%
,G_{0},\Lambda_{0}\right)  $ such that $\rho_{0}>0.$ As we can see, it is
verified the relationship $\frac{G\rho}{c^{2}}=t^{-2}.$

Therefore, at this time we have the following behavior for $G(t)$
\begin{equation}
G(t)=G_{0}t^{2(c_{1}-1-\frac{b}{2a})},\qquad G\thickapprox\left\{
\begin{array}
[c]{l}%
\text{decreasing if }\left(  c_{1}-1-b/2a\right)  <0,\\
\text{constant if }c_{1}=1+b/2a,\\
\text{growing if }\left(  c_{1}-1-b/2a\right)  >0,
\end{array}
\right.  .
\end{equation}
while $\Lambda$ behaves as follows:%
\begin{equation}
\Lambda=\Lambda_{0}t^{-2(c_{1}+1)},,\qquad\Lambda\thickapprox\left\{
\begin{array}
[c]{l}%
\text{decreasing if }c_{1}>-1,\\
\text{constant if }c_{1}=-1,\\
\text{growing if }c_{1}<-1,
\end{array}
\right.  ,
\end{equation}
therefore $\left(  c_{1}+1\right)  >0\Longrightarrow c_{1}\in\left(
-1,\infty\right)  .$ But we have not any information about the sign of
$\Lambda_{0},$ i.e. we do not obtain more information following this way.

With regard to $H$ we find that
\begin{equation}
R=R_{0}\rho^{-1/3\left(  1+\omega\right)  }=R_{0}t^{-b/3a\left(
1+\omega\right)  },\qquad XYZ=R_{0}t^{-b/a\left(  1+\omega\right)  }.
\end{equation}
If we assume that the functions $\left(  X,Y,Z\right)  $ follow a power law
(i.e. $X=X_{0}t^{\alpha_{1}})$ then we get the following result, $Kt^{\alpha
}=R_{0}t^{-b/a\left(  1+\omega\right)  },$then, $\sum_{i}^{3}\alpha_{i}%
=\alpha=-\frac{b}{^{a\left(  1+\omega\right)  }},$ so we arrive to the same
conclusion as in section (\ref{Lie}).

The shear is calculated as follows$,$ $\sigma^{2}=\sigma_{0}^{2}%
t^{-2(c_{1}+1)},$with%
\begin{equation}
\sigma_{0}^{2}=\frac{1}{3\left(  1+\omega\right)  ^{2}}\frac{b^{2}}{a^{2}%
}+\frac{2\left(  b^{2}+ab\left(  1+\omega\right)  (c_{1}+1)+3c_{0}^{2}%
c_{1}\Lambda_{0}a^{2}\left(  1+\omega\right)  ^{2}\right)  }{3a\left(
1+\omega\right)  ^{2}\left(  a\left(  2c_{1}+1+\omega\right)  +b\right)
}-\Lambda_{0}c_{0}^{2}. \label{shear0}%
\end{equation}

To calculate the value of constants $\left(  \alpha_{i}\right)  _{i=1}^{3},$we
arrive to the same system of equations i.e.
\begin{align}
\alpha_{1}\alpha_{2}+\alpha_{1}\alpha_{3}+\alpha_{2}\alpha_{3}  &  =8\pi
\frac{G_{0}}{c_{0}^{2}}\rho_{0}+\Lambda_{0}c_{0}^{2},\\
\alpha_{2}\left(  \alpha_{2}-1\right)  +\alpha_{3}\left(  \alpha_{3}-1\right)
+\alpha_{3}\alpha_{2}-c_{1}\left(  \alpha_{3}+\alpha_{2}\right)   &
=-8\pi\frac{G_{0}}{c_{0}^{2}}\omega\rho_{0}+\Lambda_{0}c_{0}^{2},\\
\alpha_{1}\left(  \alpha_{1}-1\right)  +\alpha_{3}\left(  \alpha_{3}-1\right)
+\alpha_{3}\alpha_{1}-c_{1}\left(  \alpha_{1}+\alpha_{3}\right)   &
=-8\pi\frac{G_{0}}{c_{0}^{2}}\omega\rho_{0}+\Lambda_{0}c_{0}^{2},\\
\alpha_{2}\left(  \alpha_{2}-1\right)  +\alpha_{1}\left(  \alpha_{1}-1\right)
+\alpha_{1}\alpha_{2}-c_{1}\left(  \alpha_{1}+\alpha_{2}\right)   &
=-8\pi\frac{G_{0}}{c_{0}^{2}}\omega\rho_{0}+\Lambda_{0}c_{0}^{2},
\end{align}
which solution is:%
\begin{equation}
\alpha_{1}=\alpha_{2}=\alpha_{3}=\sqrt{\frac{8\pi G_{0}\rho_{0}}{3c_{0}^{2}%
}+\frac{\Lambda_{0}c_{0}^{2}}{3}},\qquad c_{1}=-1+\frac{4\pi G_{0}\rho
_{0}(1+\omega)}{c_{0}\sqrt{\frac{8\pi G_{0}\rho_{0}}{3}+\frac{\Lambda_{0}%
c_{0}^{4}}{3}}},
\end{equation}
finding again that this kind of solutions lacks of any interest.

\section{Study of eq. (29).\label{appB}}

The purpose of this appendix is to show that if we consider the third order
ODE obtained from the field equations but without taking into account the
effects of a $c-$var into the curvature tensor we arrive to the same results
as in the above appendix. i.e. appendix (\ref{appA}). Therefore we reproduce
again the same steeps as in section (\ref{Lie}) as well as in appendix
(\ref{appA}) in order to find the restrictions for $G(t),c(t)$ and
$\Lambda(t)$ for which our field equations admit symmetries i.e. they are integrable.

Therefore, our aim is study the following eq.
\begin{equation}
\dddot{\rho}=K_{1}\ddot{\rho}\frac{\dot{\rho}}{\rho}-K_{2}\frac{\dot{\rho}%
^{3}}{\rho^{2}}+\frac{G\rho^{2}}{c^{2}}\left[  K_{3}\frac{G^{\prime}}{G}%
-K_{4}\frac{\rho^{\prime}}{\rho}-K_{5}\frac{c^{\prime}}{c}\right]  -K_{6}\rho
cc^{\prime}\Lambda, \label{Bdefeq}%
\end{equation}
where the constants $\left(  K_{i}\right)  _{i=1}^{6}$ are given by eqs.
(\ref{choren1}). Following the standard procedure we need to solve the next
system of PDEs:%
\begin{align}
c^{4}\rho^{3}\xi_{\rho}  &  =0,\label{Blie1}\\
c^{4}\rho^{3}\xi_{\rho\rho}  &  =0,\label{Blie2}\\
K_{1}c^{4}\rho\eta-K_{1}c^{4}\rho^{2}\eta_{\rho}-9c^{4}\rho^{3}\xi_{t\rho
}+3c^{4}\rho^{3}\eta_{\rho\rho}  &  =0,\label{Blie3}\\
-K_{1}c^{4}\rho^{2}\eta_{t}+3c^{4}\rho^{3}\eta_{t\rho}-3c^{4}\rho^{3}\xi_{tt}
&  =0,\label{Blie4}\\
K_{1}c^{4}\rho^{2}\xi_{\rho\rho}+K_{2}c^{4}\rho\xi_{\rho}-c^{4}\rho^{3}%
\xi_{\rho\rho\rho}  &  =0,\label{Blie5}\\
2K_{2}c^{4}\rho\eta_{\rho}-K_{1}c^{4}\rho^{2}\eta_{\rho\rho}+2K_{1}c^{4}%
\rho^{2}\xi_{t\rho}-2K_{2}c^{4}\eta-3c^{4}\rho^{3}\xi_{t\rho\rho}+c^{4}%
\rho^{3}\eta_{\rho\rho\rho}  &  =0,\label{Blie6}\\
3K_{2}c^{4}\rho\eta_{t}-2K_{1}c^{4}\rho^{2}\eta_{t\rho}+3K_{4}c^{2}\rho
^{4}G\xi_{\rho}+K_{1}c^{4}\rho^{2}\xi_{tt}-3c^{4}\rho^{3}\xi_{tt\rho}%
+3c^{4}\rho^{3}\eta_{t\rho\rho}  &  =0, \label{Blie7}%
\end{align}%
\[
3c^{4}\rho^{3}\eta_{tt\rho}-c^{4}\rho^{3}\xi_{tt\,t}+K_{4}c^{2}\rho
^{4}G\left(  \xi\left(  \frac{G^{\prime}}{G}-2\frac{c^{\prime}}{c}\right)
+\frac{\eta}{\rho}+2\xi_{t}\right)  -
\]%
\begin{equation}
-K_{1}c^{4}\rho^{2}\eta_{tt}-4K_{3}c^{2}\rho^{5}G^{\prime}\xi_{\rho}%
+4K_{5}cc^{\prime}\rho^{5}G\xi_{\rho}+4K_{6}c^{5}c^{\prime}\rho^{4}\Lambda
\xi_{\rho}=0, \label{Blie8}%
\end{equation}
\[
c^{4}\rho^{3}\eta_{tt\,t}+K_{4}c^{2}\rho^{4}G\eta_{t}+
\]%
\[
c^{2}\rho^{5}GK_{3}\left(  \xi\left(  2\frac{c^{\prime}}{c}\frac{G^{\prime}%
}{G}-\frac{G^{\prime\prime}}{G}\right)  +\frac{G^{\prime}}{G}\left(
\eta_{\rho}-2\frac{\eta}{\rho}-3_{t}\right)  \right)  +
\]%
\[
c^{2}\rho^{5}GK_{5}\left(  \xi\left(  \frac{c^{\prime}}{c}\frac{G^{\prime}}%
{G}-3\frac{(c^{\prime})^{2}}{c^{2}}+\frac{c^{\prime\prime}}{c}\right)
+\frac{c^{\prime}}{c}\left(  2\frac{\eta}{\rho}-\eta_{\rho}+3\xi_{t}\right)
\right)  +
\]%
\begin{equation}
K_{6}c^{6}\rho^{4}\Lambda\left[  \xi\left(  \frac{(c^{\prime})^{2}}{c^{2}%
}+\frac{c^{\prime\prime}}{c}+\frac{c^{\prime}}{c}\frac{\Lambda^{\prime}%
}{\Lambda}\right)  +\frac{c^{\prime}}{c}\left(  \frac{\eta}{\rho}-\eta_{\rho
}+3\xi_{t}\right)  \right]  =0 \label{Blie9}%
\end{equation}

Imposing the symmetry $X=\left(  at+e\right)  \partial_{t}+b\rho\partial
_{\rho},$ i.e. $\left(  \xi=at+e,\,\eta=b\rho\right)  ,$ ,we get the following
restrictions for $G(t),c(t)$ and $\Lambda(t)$.

From eq. (\ref{Blie8}) we get
\begin{equation}
\frac{G^{\prime}}{G}-2\frac{c^{\prime}}{c}=-\frac{b+2a}{at+e}, \label{Brest1}%
\end{equation}
while from eq. (\ref{Blie9}) it is obtained:%
\begin{align}
\left(  at+e\right)  \left(  2\frac{c^{\prime}}{c}\frac{G^{\prime}}{G}%
-\frac{G^{\prime\prime}}{G}\right)  -\left(  b+3a\right)  \frac{G^{\prime}%
}{G}  &  =0,\label{Brest2}\\
\left(  at+e\right)  \left(  \frac{c^{\prime}}{c}\frac{G^{\prime}}{G}%
-3\frac{(c^{\prime})^{2}}{c^{2}}+\frac{c^{\prime\prime}}{c}\right)
+\frac{c^{\prime}}{c}\left(  b+3a\right)   &  =0,\label{Brest3}\\
\left(  at+e\right)  \left(  \frac{(c^{\prime})^{2}}{c^{2}}+\frac
{c^{\prime\prime}}{c}+\frac{c^{\prime}}{c}\frac{\Lambda^{\prime}}{\Lambda
}\right)  +3a\frac{c^{\prime}}{c}  &  =0, \label{Brest4}%
\end{align}
where $a,b,e\in\mathbb{R}.$ Note that $\left[  a\right]  =\left[  b\right]
=1,$ i.e. they are dimensionless constants but $\left[  e\right]  =T$, with
respect to a dimensional base $B=\left\{  L,M,T\right\}  .$

Now from (\ref{Brest3}) we get:%
\begin{equation}
\frac{G^{\prime}}{G}-3\frac{c^{\prime}}{c}+\frac{c^{\prime\prime}}{c^{\prime}%
}=-\frac{3a+b}{at+e}, \label{Brest3-1}%
\end{equation}
and taking into account eq. (\ref{Brest1}) we find that%
\begin{equation}
\frac{c^{\prime\prime}}{c^{\prime}}-\frac{c^{\prime}}{c}=-\frac{a}{at+e}.
\label{Brest3-2}%
\end{equation}

In the same way, form eq. (\ref{Brest4}) it is found that
\begin{equation}
\frac{c^{\prime}}{c}+\frac{c^{\prime\prime}}{c^{\prime}}+\frac{\Lambda
^{\prime}}{\Lambda}=-\frac{3a}{at+e}, \label{Brest4-1}%
\end{equation}
and taking into account eq., (\ref{Brest3-2}) we get:%
\begin{equation}
\frac{\Lambda^{\prime}}{\Lambda}+2\frac{c^{\prime}}{c}=-\frac{2a}{at+e}.
\label{Brest4-2}%
\end{equation}

These restrictions will be enough to find a solution for eq. (\ref{Bdefeq})
i.e. eqs. (\ref{Brest3-2}, \ref{Brest1} and \ref{Brest4-2}). As it is observed
we have arrive to the same restrictions as in section (\ref{Lie}) as well as
in appendix (\ref{appA}). Therefore, following this approach, there is no
difference between to consider $c$-var affecting to the curvature tensor and
to consider the usual FE. We will show that we arrive to the same result in
the case of the scale symmetry, the other solutions are obtained in the same
way following the steeps as in section (\ref{Lie}).

\subsection{Scale symmetry.}

Making $e=0$ $i.e.$ considering only $\left(  \xi=at,\quad\eta=b\rho\right)
,$ we have to integrate eqs. (\ref{Brest3-2}, \ref{Brest1} and \ref{Brest4-2}%
), so
\begin{align}
\frac{c^{\prime\prime}}{c^{\prime}}-\frac{c^{\prime}}{c}  &  =-\frac{1}{t},\\
\frac{G^{\prime}}{G}-2\frac{c^{\prime}}{c}  &  =-\frac{b+2a}{at}%
,\qquad\Longrightarrow\qquad\frac{G}{c^{2}}=Bt^{-\left(  2+\frac{b}{a}\right)
}\\
\frac{\Lambda^{\prime}}{\Lambda}+2\frac{c^{\prime}}{c}  &  =-\frac{2}%
{t},\qquad\Longrightarrow\qquad\Lambda c^{2}=\tilde{B}t^{-2},
\end{align}
$B,\tilde{B}\in\mathbb{R},$ therefore we get
\begin{align}
c  &  =c_{0}t^{c_{1}},\qquad c_{1},c_{0}\in\mathbb{R},\label{Bcscale}\\
G  &  =G_{0}t^{2(c_{1}-1-\frac{b}{a})},\qquad G_{0}\in\mathbb{R}%
^{+},\label{BGscale}\\
\Lambda &  =\Lambda_{0}t^{-2(c_{1}+1)},\qquad\Lambda_{0}\in\mathbb{R},
\label{BLscale}%
\end{align}
where we assume that $G_{0}>0.$ The invariant solution for the energy density
is: $\rho=\rho_{0}t^{b/a},$ and for physical reasons we impose the condition,
$ab<0$ then $b<0.$ If we make that this solution verifies eq. (\ref{Bdefeq})
with $c(t),G(t)$ and $\Lambda(t)$ given by eqs. (\ref{Bcscale}-\ref{BLscale}),
we find the value of constant $\rho_{0},$ so%
\begin{equation}
\rho_{0}=-\left[  \frac{c_{0}^{2}\left(  b^{2}+ab\left(  1+\omega\right)
+3c_{0}^{2}c_{1}\Lambda_{0}a^{2}\left(  1+\omega\right)  ^{2}\right)  }{12\pi
aG_{0}\left(  1+\omega\right)  ^{2}\left(  a\left(  2c_{1}+1+\omega\right)
+b\right)  }\right]  ,
\end{equation}
with the only restriction $\omega\neq-1,$ compare with eqs. (\ref{ayla} and
\ref{ayla1})$.$ Note that $ab<0,$ so we need to choice constants $\left(
c_{1},c_{0},G_{0},\Lambda_{0}\right)  $ such that $\rho_{0}>0.$ As we can see,
it is verified the relationship, $G\rho/c^{2}=t^{-2},$ i.e. the Mach
relationship for the inertia. Therefore, at this time we have the following
behavior for $G(t)$
\begin{equation}
G(t)=G_{0}t^{2(c_{1}-1-\frac{b}{2a})},\qquad G\thickapprox\left\{
\begin{array}
[c]{l}%
\text{decreasing if }\left(  c_{1}-1-b/2a\right)  <0,\\
\text{constant if }c_{1}=1+b/2a,\\
\text{growing if }\left(  c_{1}-1-b/2a\right)  >0,
\end{array}
\right.  .
\end{equation}
while $\Lambda$ behaves as follows:%
\begin{equation}
\Lambda=\Lambda_{0}t^{-2(c_{1}+1)},,\qquad\Lambda\thickapprox\left\{
\begin{array}
[c]{l}%
\text{decreasing if }c_{1}>-1,\\
\text{constant if }c_{1}=-1,\\
\text{growing if }c_{1}<-1,
\end{array}
\right.  ,
\end{equation}
therefore $\left(  c_{1}+1\right)  >0\Longrightarrow c_{1}.\in\left(
-1,\infty\right)  .$ But we have not any information about the sign of
$.\Lambda_{0}.$ If we assume that the functions $\left(  X,Y,Z\right)  $
follow a power law (i.e. $X=X_{0}t^{\alpha_{1}})$ then we get the following
result, $Kt^{\alpha}=R_{0}t^{-b/a\left(  1+\omega\right)  },$ and $\sum
_{i}^{3}\alpha_{i}=\alpha=-\frac{b}{^{a\left(  1+\omega\right)  }}.$ The shear
is calculated as follows$,$ $\sigma^{2}=\sigma_{0}^{2}t^{-2}.$

So as we can see it is obtained the same solution, with the same order of
magnitude and therefore we conclude that there is no difference between both approaches.

\end{document}